\documentclass[twocolumn,showpacs,preprintnumbers,amsmath,amssymb]{revtex4-1}

\usepackage[dvips]{graphicx}
\usepackage[dvipdfmx]{hyperref}
\usepackage{dcolumn}
\usepackage{bm}

\newcommand{\braket}[1]{\langle\,{#1}\rangle}
\def\nn{\nonumber\\}
\def\bea{\begin{eqnarray}}
\def\eea{\end{eqnarray}}
\def\ol{\overline}
\def\wt{\widetilde}
\def\wh{\widehat}
\def\pa{\partial}
  \def\R{\mathbf R}  
\def\Vec#1{\mbox{\boldmath $#1$}}
\def\x{\Vec{x}}
\def\sgn{\mathop{\rm sgn}}
\newtheorem{proposition}{Proposition}[section]

\begin{document}

\title{Bubbly vertex dynamics: a dynamical and geometrical model for epithelial tissues with curved cell shapes}
\author{Yukitaka Ishimoto}
\email{ishimoto@cdb.riken.jp}
\author{Yoshihiro Morishita}
\affiliation{
Laboratory for Developmental Morphogeometry,
RIKEN Center for Developmental Biology,
Kobe 650-0047, Japan
}

\begin{abstract}
In order to describe two-dimensionally packed cells in epithelial tissues both mathematically and physically, there have been developed several sorts of geometrical models, such as the vertex model, the finite element model, the cell-centered model, the cellular Potts model. So far, in any case, pressures have not neatly been dealt with and the curvatures of the cell boundaries have been even omitted through their approximations. We focus on these quantities and formulate them in the vertex model. Thus, a model with the curvatures is constructed, and its algorithm for simulation is provided. The possible extensions and applications of this model will also be discussed.
\end{abstract}

\pacs{87.17.Aa, 87.17.Pq, 87.17.Rt, 87.18.Fx}

\maketitle

\section{Introduction}

Cellular patterns occur frequently in nature, such as cells in biological tissues, crystalline grains in polycrystals, grain aggregates in colloidal materials, and bubbles in a pint of beer. Among cellular patterns, the dynamical behaviors of epithelial tissues have received much attention in the study of morphogenesis in developmental biology as well as in biophysics and soft matter physics \cite{levental07}. It is because they provide a firm interdisciplinary field to explore the various biological, chemical, and physical properties of the cellular materials with autonomous characters and collective behaviors. However, even the phenomena linked to the mechanical properties of tissue, such as cell boundary shortening, cell division, cell rearrangement, and cell disappearance from the tissue, are highly complex, with thousands of different molecules structured and orchestrated in a very intricate way. By contrast, it is widely recognized that the apical side of the cell-cell junction (AJ), such as the adherens junction, of monolayered epithelial tissue forms a two-dimensional elastic network, and the AJ network appears approximately as a polygonal cellular pattern on the two-dimensional surface \cite{honda83,oster90,honda01,honda04,farhadifar07,honda78,gibson09,gibson11}. In addition, an elastic, possibly visco-elastic-plastic, network containing the AJ network likely plays an essential role in the mechanical and dynamical properties of cell movement in the epithelial tissues. Therefore, dynamical and geometrical models would describe certain aspects of various phenomena in epithelial tissues.

\begin{figure}[h]
  \centering
  \includegraphics[width=4.75cm]{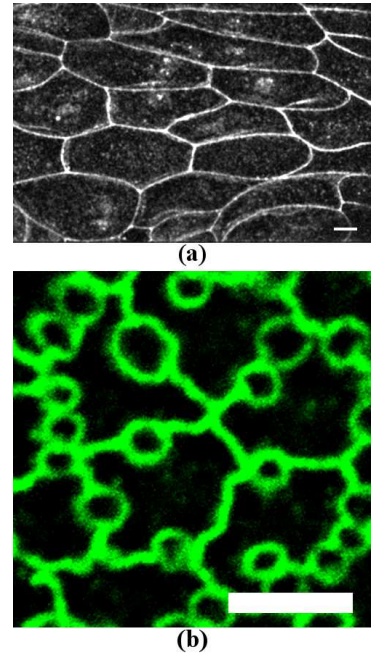}
  \caption{Examples of epithelial tissue with curved cell shapes. (a) Epithelial tissue in the abdomen of {\it Drosophila} pupa, 19hr15min APF, DECadGFP knock-in \cite{huang09}, courtesy of K. Sugimura (WPI-iCeMS, Kyoto Univ.). Scale bar: 10 $\mu$m. (b)  Apical surface view of the olfactory epithelium. Olfactory epithelium consists of two different types of cells, olfactory cells (small, rounded shape) and supporting cells. Olfactory epithelium of mouse was stained with ZO-1. Courtesy of S. Katsunuma and H. Togashi (Graduate School of Medicine, Kobe Univ.). Scale bar: 5 $\mu$m.}
  \label{fig:flywing}
  \label{fig:olfactory}
\end{figure}

Experimental observations indicate that some cell boundaries of epithelial tissue have smooth nonzero curvatures that cannot be well approximated by a polygonal pattern with straight boundaries \cite{stewart11,gibson06,gibson11,lecuit07,togashi11} (Fig. \ref{fig:flywing}). 
For example, mitotic cell rounding \cite{stewart11,gibson06,gibson11}, retinal epithelium of fruit fly \cite{lecuit07}, two-vertex cells of mouse olfactory epithelium (Fig. \ref{fig:olfactory}(b)), and swollen cell species in other epithelial tissues \cite{togashi11}, require precise treatment of the curvatures for their precise descriptions. In HeLa cells \cite{stewart11}, cortical tension and pressures generally increase by substantial amounts during mitotic cell rounding and play important roles in the phenomenon. Therefore, a dynamical and geometrical model with curvatures may provide a firm theoretical background to explore such various morphologies and dynamical behaviors in epithelial tissues. An optimal model would feature better precision for curvatures and cell internal pressure and be extendable to various biochemical interactions for its applications.

In order to theoretically describe two-dimensionally packed cells in epithelial tissues, there have been developed several types of geometrical models \cite{gibson09}, such as the vertex model \cite{honda83,nagai88,oster90,kawasaki89,fuchizaki95,honda01,honda04,farhadifar07,farhadifar10,fletcher13}, the finite element model \cite{brodland00}, the cell-centered model \cite{honda83,honda78,ciudad10}, and the cellular Potts model (an extended Potts model) \cite{graner92,glazier93}. Among these models, the vertex model is the first and simplest to be generalized with curvatures. The finite element model partitions polygonal cells into pieces so that the cell boundaries are likely to be piece-wised but otherwise behaves like the vertex model. The cell-centered model is based on the Voronoi diagram created from the centers of the cells, whose boundaries are secondary objects and are mostly ignored. The cellular Potts model is defined on a fine lattice in which the shapes of the cell boundaries are, in general, arbitrarily pixelated by definition. Thus, the latter three families of models do not fit naively with the simplest view of smoothly curved cell boundaries. The vertex model for cells was first proposed by Honda \cite{honda83} and is a mechanical network model of tricellular junctions in which the junctions are simplified to ``fictitious'' vertices. Therefore, if one constructs a model of cell boundaries with curvatures, which we intend to construct here, it can be easily merged with the vertex model. Note that a straightforward simplification of biological imaging data is to assign the data to the cell boundaries and the individual cells, which matches the parameters of the vertex model quite naively. In addition, some integral membrane proteins bind tricellular junctions, forming the tricellular tight junctions \cite{ikenouchi05}. The formation of these junctions might support the validity of the ``fictitious'' vertices in the vertex model.

Of relevance to the problem of modeling the curvatures of cell boundaries is the Young-Laplace law \cite{younglaplace}, which is a fundamental law in the physics of densely packed dry foams \cite{weaire99}. The law represents the force balance along the cell boundary and equates the pressure difference between the neighboring cells and the tension of the boundary in its normal direction. The law is applicable to any kind of cell boundaries of the cellular materials if such forces are acting on them and their bending rigidity is negligible relative to the size of the cell boundaries. Due to the thinness of the plasma membrane and the flexibility of the cortical actin fibers, it would also be legitimate in our case. The law dictates the behaviors and the accuracy of the pressures are essential for the curvatures and vice versa, if the pressures depend on the curvatures through the cell areas.

Previous studies have not adequately addressed pressures, and curvatures have even been omitted through the polygonal approximations of the geometrical models for epithelial tissues. Apart from the cellular Potts model, which we exclude from our current consideration of the curvature model, no study has attempted to capture non-straight cell boundaries in geometrical models for epithelial tissues, and studies of simpler cellular materials for general purposes have been limited \cite{kawasaki89,fuchizaki95,brakke92}. In a straightforward and computer-friendly approach, ``virtual'' vertices were inserted in the interior of the cell boundaries, which may be classified as a version of the finite element method. A superior example of this approach is the computer program Surface Evolver \cite{brakke92}, which is a very general and powerful tool to find the minimal surface in arbitrary dimensions that can even be equipped with an arbitrary Riemannian metric. In contrast to its generality, the Surface Evolver minimizes the surface energy rather than simulating its dynamics. In another attempt, similar ``virtual'' vertices were introduced with the associated local curvatures. Such a model has been constructed by reducing the dissipative equation of motion for two-dimensional grain growth \cite{kawasaki89}, and for three-dimensional cellular pattern growth \cite{fuchizaki95}. The local curvatures appear in the potential force for the virtual vertices and shift them according to the boundary tension. Both attempts are strong candidates for application to epithelial tissues, however, if the dynamics in question fits to their scopes. 

During development, for example, the dynamics of epithelial tissues feature slow movements of the cells in hours and days, and even during the rapid change in curvature that occurs during mitosis, cell divisions are performed in approximately half an hour. By contrast, laser ablation experiments of cell boundaries imply that the initial forces along the cell boundaries can be released within a second \cite{farhadifar07,hutson09,bonnet12,ishihara12,sugimura13,mao13}. This rapid release indicates that the curvatures are the ``fast'' variables while the cellular positions are the ``slow'' variables. The dynamical degrees of freedom for the curvatures are effectively suppressed for the cellular dynamics of the epithelial sheet. Therefore, the optimal means of addressing cellular dynamics is to treat the curvatures mostly in a static way and to increase the accuracy of the geometrical description of the tissues via the Young-Laplace law, which we will attempt here.

In this paper, we aim to construct a two-dimensional geometrical and dynamical model of the univariant curvatures of cell boundaries and the tricellular junctions to serve as a tool for investigating the various dynamical phenomena of epithelial tissues with greater accuracy. The degrees of freedom of the variables are, therefore, equal to the number of the boundaries and twice that of the vertex positions. To apply the Young-Laplace law, a general form of the potential energy is proposed as a function of the curvatures and the vertex positions. The generalization also offers a clear physical interpretation of the energy and a variety of mechanical responses of the cells, which will be discussed in detail in Section \ref{sec:target area} and \ref{sec:discussions}. From the energy, cell pressures and tensions are defined in a physically and mathematically sound manner, and appear in the law as a consequence of the ``fast'' curvature variables. Thereafter, the dynamics is formulated by the dissipative equations of motion for the vertex positions with some topological changes called T1 and T2 \cite{weaire84}, while the curvatures are determined by the law. Because the law emerges as a set of non-linear equations, we present a numerical algorithm with the method of steepest descent for the determination of the curvatures. 

During the construction of the model, the buckling effects of the cell boundaries are neglected, and we introduce the artifact of the upper and lower bounds of the curvatures to avoid further and sometimes unnatural complications in simulation. If we allow an extreme initial configuration and infinitely large circular cell boundaries, the model admits many overlapping boundaries; otherwise, boundary collisions should be treated carefully with the buckling effects. The latter requires a precise treatment of the elastic nature of the cell areas and the nonuniform curvatures, which would require elliptic integrals to express. These considerably complicate the construction and are beyond the scope of this manuscript. The form of the energy contains linear surface tensions and osmotic pressures in addition to the conventional terms in the vertex models, and thus the area part of the energy may permit a longer relaxation time of the largely deformed area compared to the conventional one. This form might match the slow movement of the deformed cells around the healed wound mentioned in \cite{honda01}. We omit angular energies in the form of the energy due to a lack of its clear physical correspondence. 

The model enables us to describe the dynamics of two-dimensional epithelial sheets with rounded cell shapes. Most notably, the model further admits two-vertex or two-sided cells (Fig. \ref{fig:olfactory}(b)) to appear, which cannot even exist in the conventional vertex models. We demonstrate such dynamics by simulation in Section \ref{sec:examples}, and give a brief comparison with an experimental observation. The simulation is performed for a few hundred packed cells with a boundary with the free boundary condition. We also provide further discussions and comments on the model, its possible extensions, and its applications in the final section.

\section{Model}

This section consists of three parts: the preliminary part, the construction of the dynamical model, which we call the bubbly vertex dynamics, and some examples of simulation based on the model.

\subsection{Preliminaries}
\subsubsection{The Young-Laplace law and the tricellular junction}

In \cite{younglaplace}, the Young-Laplace law or equation is formulated in the context of physics to explain the equilibrium shape of the interface between two media that exert their capillary pressures onto the interface. Denoting the pressure difference by $\Delta p$ and the unit vector normal to the surface by $\Vec{n}$, the local form of the Young-Laplace law can be expressed as
\bea
 - T\, \nabla\cdot\Vec{n}  &=& \Delta p,
\label{def:3dYL}
\eea
where $T$ is the positive tension along the interface and $\nabla$ is the differential operator on the coordinate system of the interface. Thus, if the interface is two-dimensional, the derivative along $\Vec{n}$ contributes nothing, and $\nabla\cdot\Vec{n}$ amounts to twice the mean curvature of the interface. The law is frequently used in models of grain growth and soap froths \cite{weaire99}. An important remark on Eq. (\ref{def:3dYL}) is that positive tension is assumed because otherwise the interface is buckled or even shuffled if the two media are in liquid phase.

\begin{figure}[h]
  \centering
  \includegraphics[width=4cm]{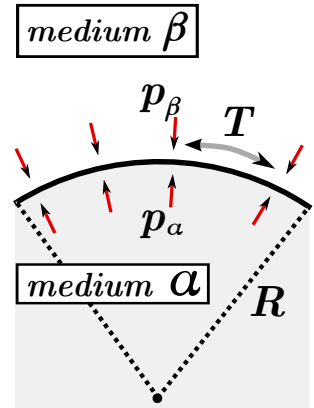}
  \caption{An illustration for the Young-Laplace law in two dimensions. A uniform tension $T$ along the one-dimensional interface with the curvature radius $R$ is balanced by the pressure difference $p_\alpha-p_\beta$.}
  \label{fig:YL}
\end{figure}

As illustrated in Fig. \ref{fig:YL}, in two dimensions with a uniform $T$ along the one-dimensional interface, the above can be simplified to 
\bea
  \frac{T}{R} &=& p_\alpha - p_\beta.
\eea
We specify the pressures of the media $\alpha$ and $\beta$ by $p_\alpha$ and $p_\beta$, respectively. $R$ denotes the radius of the uniform curvature and is positive, assuming $p_\alpha>p_\beta$. If we consider a tricellular junction at its force equilibrium, there arise three equations from the law:
\bea
  \frac{T_i}{R_i} &=& p_i - p_{i+1},
\eea
where adjacent cell boundaries are labeled as $i=1,2,3$ and the $i$-th boundary partitions the region $i$ and $i+1$ with the identification: region $4$ $\sim$ region $1$. Summing the above three equations yields
\bea
  \sum_{i=1}^{3} \frac{T_i}{R_i} = 0.
  \label{eq:triYL}
\eea
Therefore, at the force equilibrium, there is a relationship between the tensions and the curvatures, in addition to the direct force balance equation at the junction. Although we do not intend to utilize the above in the rest of the manuscript, this would be a very useful and important relationship, for instance, when one wants to infer the forces from images \cite{ishihara12,ishihara13}. Note that an equivalent equation to Eq. (\ref{eq:triYL}) in the case of foams is given in \cite{durand10}, and that the law is also mentioned in the context of vertex models in \cite{chiou12}.

\subsubsection{Vertex Dynamics}

The vertex model for cells in tissues was first proposed by Honda in his seminal paper \cite{honda83} as the boundary shortening model in two dimensions, which finds the minimum of the total boundary length for a given network of vertices. The model has subsequently been developed with various terms and formulations that fall roughly into two families. One includes equations of motion for vertices focusing on its dynamics, while the other deals with the ensemble of the network configurations energetically, for example, by Monte Carlo simulation. The former is sometimes referred to as the ``vertex dynamics'', while the latter is simply referred to as the vertex model. Rather than exhaustively listing these models, we mention some of them below. Nagai et al. formulated the vertex dynamics with a set of deterministic and dissipative equations of motion \cite{nagai88, kawasaki89}. Weliky and Oster developed it with the two-dimensional osmotic pressure and the Hookean spring term for the cell perimeter \cite{oster90}. Various extensions to three-dimensional cases have also been presented \cite{honda01, fuchizaki95, honda04, okuda13}. Farhadifar et al. and Staple et al. basically used the latter energetic treatment to define the model and performed ground-state analysis for an extensive set of model parameters \cite{farhadifar07,farhadifar10}. Marinari et al. adapted Monte Carlo simulations of the vertices for tissue overcrowding \cite{marinari12}. In either family of the models, there are mainly three building blocks: the polygonal approximation of a given cellular pattern, the potential energy function to characterize the tissue, and the transition rules for the network configurations, including the dynamical equations and computer algorithms.

The polygonal approximation lets the geometrical features of the tissue be a graph of vertices and edges in which vertices represent the multicellular junctions and edges represent the cell boundaries. The edges are defined as straight lines, and in most cases, each vertex is assumed to be connected only to three edges (the law of Plateau in dry foams \cite{weaire99}). We call it the degree of the vertex is three. Let us label the two-dimensional position of the $i$-th vertex by $\Vec{x}_i$. The original potential energy proposed in \cite{honda83} is merely the sum of the cell boundary lengths. It has been subsequently developed and extended to various directions, and one of the most accepted and frequently used forms of the energy for epithelial tissues is that described in \cite{farhadifar07,farhadifar10}: 
\bea
  \label{def:Evdm}
  E(\{\x_i\}) = \sum_{<i,j>} \Lambda_{ij} l_{ij}
   + \sum_{\alpha} \frac{\Gamma_\alpha}{2} L_\alpha^2
   + \sum_{\alpha} \frac{\kappa_\alpha}{2} 
     \left( A_\alpha - A_\alpha^0 \right)^2 , 
\nn
\eea
where $i,j$ are vertices and $\alpha$ specifies a cell. $\Lambda_{ij}$ is a constant, and $\Gamma_\alpha$ and $\kappa_\alpha$ are non-negative constants. $l_{ij}$ represents the length of the cell boundary and, consequently, the distance between the $i$-th and the $j$-th vertices. $A_\alpha$ is the area of the cell $\alpha$, while $A_\alpha^0$ is its target area, or the preferred area. The first summation is over the connected neighbors. The third term, ``area constraint'', is introduced as ``a mechanism that makes the size of the cells uniform'' in \cite{honda01}.

The equations of motion for the vertex positions are given by
\bea
  \eta_i \frac{d \x_i}{dt} = - \frac{\pa E (\{\x_i\})}{\pa \x_i}.
\eea
For a given initial configuration of $\{\Vec{x}_i\}$, the vertices move dissipatively, thereby obeying the above equations. Certain topological changes of the vertex graph are observed in epithelial tissues. The main contributions are the recombination of the vertices and the cell disappearance from the surface. We discuss their details in Section \ref{sec:topological changes}

As is often the case with the vertex models, we assume the tensions along the cell boundaries are all non-negative in the following sections. Our construction seems to have a room to admit negative tensions, but non-uniform curvatures should be introduced to describe the buckling, which we neglect in the manuscript.

\subsection{Bubbly Vertex Dynamics}
\label{sec:BVD}

This section is devoted to the construction of our model and describes the energy function, the target area, the energy minimization, the equations of motion, the determination of the curvatures by the Young-Laplace law, and the topological changes to be completed. Some technical notes and proofs of our claims are provided in the appendices. Before going into the construction, we would like to introduce four assumptions, three of which are quite often implied in the vertex models.

First, the degrees of the vertices are all three, unless otherwise stated. In the manuscript, we omit the vertices with degree two which can represent the corners of the tissue boundaries, although these vertices can be added in a straightforward manner. As for the vertices with degree greater than three, they are of importance as the multicellular rosette structure \cite{blankenship06}. However, in practice, they cannot clearly be distinguished from a collection of vertices with degree three connected by short edges. Therefore, it is assumed that these vertices can be well approximated by the vertices with degree three, and these vertices are thus be neglected. As suggested in \cite{farhadifar10}, vertices with degree greater than three are energetically unstable in many cases due to the state of the infinitesimally separated vertices with degree three, which lowers the energy. This situation is also implied in the assumption.

Second, non-negative tension is assumed on the cell boundaries so that no buckling occurs. Third, the energy function is composed of the edge lengths and the cell areas only. We simply neglect other possibilities such as boundary bending energies, angular energies, and chemical potentials. The buckling effect and the bending stiffness require nonuniform curvatures of the boundaries, while the chemical potentials require other information than the mechanical properties of the tissues. These phenomena are beyond the current scope of constructing a dynamical and geometrical model with smooth and uniform curvatures. 

Finally, we address an additional assumption for epithelial tissues: the dynamical degrees of freedom of the curvatures are negligible as the fast variables. This assumption allows us to focus on the slow dynamics of the vertex positions. We dictate more on this in Section \ref{sec:minimisation}.

\subsubsection{The energy function}
\label{sec:energy}

Emphasizing the local pressures of cells and the curvatures of cell boundaries, let us propose the bubbly vertex dynamics (BVD) model, which is inspired by the soap froths models \cite{weaire99} and the vertex models initiated by Honda \cite{honda83}. Namely, each cell has a unique hydrostatic pressure, and each cell boundary has a unique curvature resulting from the mechanical properties of the boundary and the surrounding pressures, as in the Young-Laplace law. Because a curvature of a boundary is introduced as such, the length of the boundary is no longer the distance between two ends, but is given by the curvature. Similarly, the area of a cell depends on the curvatures surrounding it. To formulate a curvature model with a mechanical vertex network, let us first suppose a general form of the (potential) energy function and deduce the cell pressures, the Young-Laplace law, and the equations of motion of the vertices in later sections. 

Let $\{\x_i\}$ be the positions of the vertices, {\it i.e.}, the junctions of the cell boundaries, and $\{R_{ij}\}$ be the radii of the curvatures of the cell boundaries. We then define the energy as
\bea
\label{def:E}
&&
  E( \left\{ \x_i \right\}, \left\{ R_{ij} \right\} ) 
\nn&&
   = \sum_{\braket{i,j}} \Lambda_{\alpha \beta}\, l_{ij}
    + \sum_{\braket{i,j}} \frac{\Gamma_{\alpha\beta}^{(l)} }{2} (l_{ij})^2
    + \sum_{\alpha} \frac{\Gamma_{\alpha}^{(L)} }{2} L_{\alpha}^2
\nn&&\quad
    + \sum_\alpha \left\{ \kappa_\alpha^{(1)} A_\alpha + \frac{\kappa_\alpha^{(2)}}{2}  \left( A_\alpha - A_\alpha^{(0)} \right)^2 \right\}
\nn&&\quad
    - \sum_{\alpha} \int_{A_\alpha^{(0)}}^{A_\alpha} dS_\alpha \; \Pi_\alpha(S_\alpha) 
    - P_{outer} \left( \sum_\alpha A_\alpha \right),   
\eea
where $i$ and $j$ run on the vertex numbering, while $\alpha$ runs on the cell numbering. The pair $\braket{i,j}$ stands for the edge (cell boundary) connecting the vertices $i$ and $j$. $\Lambda_{\alpha\beta}$ is a constant depending on the cells, $\alpha$ and $\beta$, on both sides of the edge, as is $\Gamma_{\alpha\beta}^{l}$. $l_{ij}$ is the length of the edge, not the distance of the two end vertices $i$ and $j$. $\Gamma_\alpha^{(L)}$ is a non-negative constant for the cell $\alpha$, and $L_\alpha$ is the perimeter of the cell. $A_\alpha$ is the area of the cell $\alpha$, and $\kappa_\alpha^{(1,2)}$ are non-negative numbers exclusive to each other that may depend on the height of the cell for epithelial tissues \cite{honda01}. $A_\alpha^{(0)}$ is the so-called ``target area'' in the quadratic form of $A_\alpha$. $dS_\alpha$ is the surface element of the cell $\alpha$ for the integration of the osmotic pressure $\Pi_\alpha$, whose functional form should be given according to the conditions and the situations one considers. We will propose an explicit form of the osmotic pressure later in Eq. (\ref{def:osmo}). $P_{outer}$ is the pressure of the outer region of the tissue, if there is a boundary of the tissue. One can easily extend this to the multiple boundary case by adding corresponding similar terms to the above. 

\begin{figure}[h]
  \centering
  \includegraphics[width=6cm]{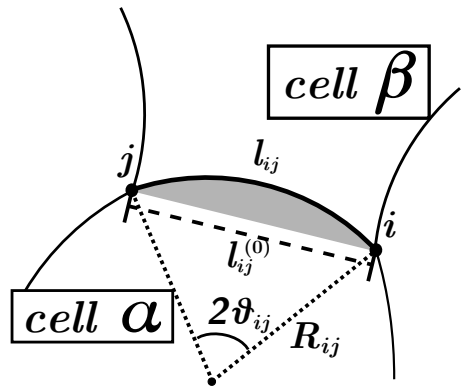}
  \caption{A simple illustration of a wedge part of the cell area.}
  \label{fig:wedge}
\end{figure}

With the above radius $R_{ij}$ of the curvature, the length $l_{ij}$ of the edge $\braket{i,j}$ and the area $A_\alpha$ of the cell $\alpha$ can be expressed by
\bea
\label{def:l}
\label{def:A}
  l_{ij} &=& 2 R_{ij} \arcsin \frac{l_{ij}^{(0)}}{2 R_{ij}},
\nn
  A_\alpha &=&
    A_\alpha^{(poly)} 
    + \sum_{\braket{i,j} \in \alpha} R_{ij}^2 \left( \theta_{ij} - \frac12 \sin 2\theta_{ij} \right) ,
\eea
where $l_{ij}^{(0)}$ denotes the distance of the two ends, $l_{ij}^{(0)} \equiv \left| \x_j - \x_i \right|$, and $A_\alpha^{(poly)}$ is the polygonal area of the cell $\alpha$ defined by the vertices. The angle $\theta_{ij}$ is depicted in Fig. \ref{fig:wedge}, satisfying $\sin \theta_{ij} = l_{ij}^{(0)}/(2 R_{ij})$, and is given by 
\bea
\label{def:theta}
  \theta_{ij} = \arcsin\left( \frac{l_{ij}^{(0)}}{2 R_{ij}} \right).
\eea
Here, we assign the numbering to the lower indices, $i$ and $j$, of $R_{ij}$ in counter-clockwise order to the cell $\alpha$, and $R_{ij}$ is positive if it is convex to the cell in question. That is, $R_{ij}$ takes a positive value when the edge is convex to the left cell of the edge $(i \to j)$. By this convention, $R_{ij} = - R_{ji}$. The definitions (\ref{def:l}, \ref{def:theta}) imply that $|\theta_{ij}| \leq \frac{\pi}{2}$, {\it i.e.}, $l_{ij}^{(0)} \leq l_{ij} \leq \frac{\pi}{2} l_{ij}^{(0)}$, so that the maximum convex (or concave) shape is a semicircular shape. We consider the curvature or the edge saturated if $|\theta_{ij}|=\frac{\pi}{2}$. One can remove this bound by using different parameterizations of the lengths and the angles, and consider larger deformations of the edges. Here, we will not consider such deformations in search of states with moderate deformations relaxing into (quasi-) equilibrium. Note that the saturation could be interpreted as a sudden increase in the bending energy or some ratchet-like mechanism of its element. Additional notes and comments on the bound will be given in Section \ref{sec:discussions}.

Let us return to the form of the energy function (\ref{def:E}). The first term of Eq. (\ref{def:E}) can be described as the ``line tension'' but we introduce this as a consequence of the combination of the cell-cell adhesion, the line tension, and the natural lengths of the Hookean springs. Namely, the second term is a Hookean spring term, assuming that an actin-filament-like elastic structure is associated with the cell boundary. The third term is also Hookean but of the ``cortical tension'' or the ``contraction ring'' \cite{oster90,farhadifar07}. The linear term in $A_\alpha$ with the coupling $\kappa_\alpha^{(1)}$ describes the surface tension, which we also assume as a two-dimensional homogeneous elastic network of the actin-filament-like structure in the apical side of the cell. The term with the coupling $\kappa_\alpha^{(2)}$ is the ``area constraint'' introduced by Nagai and Honda as the mechanism that makes the sizes of the cells uniform \cite{honda01}. As stated, the couplings $\kappa_\alpha^{(1,2)}$ are exclusive to each other because their non-zero values can be united into a single quadratic form by square completion of the linear and quadratic terms in $A_\alpha$. Although this quadratic ``effective'' term is used frequently in the literature, we will mostly omit this by setting $\kappa_\alpha^{(2)}=0$ in what follows. In fact, we suggest that this term can effectively be reproduced by the linear term in $A_\alpha$ and the osmotic pressure. This will be explained in the following section in more detail. Note that it may be possible to unify the second, third, and fourth (bracketed) terms because they all imply the existence of the elastic body on the apical side of the cell. However, we do not know such a unified form and therefore keep these terms separate. The integration of the osmotic pressure can be performed from some arbitrary but constant referential area to the value of $A_\alpha$. Later, we will actually choose a different referential area in the case of $\kappa_\alpha^{(2)}=0$. Last, the contribution of the external pressure is added, as is usual in physics.

We have not specified the explicit form of the osmotic pressure $\Pi_\alpha (A_\alpha)$ because it truly depends on the situation that the user of our model intends to realize. In addition, this two-dimensional osmotic pressure is not literally the same as the three-dimensional one. It is an isotropic force per unit length acting on the AJ boundaries and can be, for example, the swelling force of the actin gel, as indicated in \cite{oster90}, or a projection of the three-dimensional osmotic pressure or another, unspecified isotropic contribution. Here, we call it the osmotic pressure and propose the following form, which is inspired by a simple and conventional form of such pressure in chemical thermodynamics \footnote{
The more formal form of the osmotic pressure in chemical thermodynamics is $\Pi = - \frac{R T}{V} \ln(\gamma_s x_s)$, where $\gamma_s$ is the activity coefficient of the solvent, $x_s$ is the mole fraction of the solvent, and $V$ is its molar volume ($m^3/mol$). $R$ is the gas constant and $T$ is the temperature. In our two-dimensional case, $V$ is assumed to be simplified to the area multiplied by the height of the cell, and $\gamma_s x_s$ is assumed to be nearly constant.
}:
\bea
\label{def:osmo}
  \Pi_\alpha( A_\alpha )  = \frac{R_\alpha^{(o)}}{ A_\alpha + V_\alpha }, 
\eea
where $R_\alpha^{(o)}$ is some constant and the constant parameter $V_\alpha$ is introduced as an additional buffering area for practical use and to avoid the obvious singularity at $A_\alpha=0$. This expression implies that the cytoplasm is an aqueous solution containing a mixture of solutes and that the cell membrane is permeable to solvent molecules to some extent. If no solutes are transferred from one cell to another during the process, the osmotic pressure is indeed given as above with $V_\alpha=0$, the form of which coincides with the one suggested by Weliky and Oster \cite{oster90}. Note that from an energetic or dynamical point of view, we only need to specify the dependence of the pressures on the curvatures. In fact, the above specifies it as a function of the area and therefore as a composite function of the curvatures in the given way. 

\subsubsection{The target area and the simplification of the energy for homogeneous cases}
\label{sec:target area}

In this section, we focus on the area part of the energy and define the target area such that it gives the minimum of the area part. During this derivation, one can confirm that the combination of the surface energy and the osmotic pressure gives the same analytic behavior around the minimum as that given by the ``area constraint''. In addition, these two terms provide clear physical interpretations and a variety of mechanical responses of the cells. We also show further reductions of the expression (\ref{def:E}) for particular cases.

Let us consider the case without boundary for simplicity.
The energy (\ref{def:E}) can be decomposed into two parts:
\bea
\label{def:EaEl}
  E &=& E_A + E_{l},
\eea
where
\bea
\label{def:Ea}
  E_{A} &=& 
    \sum_\alpha \Bigl\{ \kappa_\alpha^{(1)} A_\alpha + \frac{\kappa_\alpha^{(2)}}{2}  \left( A_\alpha - A_\alpha^{(0)} \right)^2 
\nn&&
    - \int_{}^{A_\alpha} dS_\alpha \; \Pi_\alpha(S_\alpha) \Bigr\},
\\
  E_{l} &=& \sum_{\braket{i,j}} \Lambda_{\alpha \beta}\, l_{ij}
    + \sum_{\braket{i,j}} \frac{\Gamma_{\alpha\beta}^{(l)} }{2} (l_{ij})^2
    + \sum_{\alpha} \frac{\Gamma_{\alpha}^{(L)} }{2} L_{\alpha}^2 . 
\eea
It should be remarked here that, in the case with a boundary, the change of the vertex positions and, therefore, of the areas in the bulk has no affect on the last term of Eq. (\ref{def:E}) with $P_{outer}$. Only the cells at the boundary contribute to the change in the boundary term. Conversely, the boundary term is relevant only for the dynamics at the boundary.

\begin{figure}[h]
\vspace{10pt}
  \centering
  \includegraphics[width=7cm]{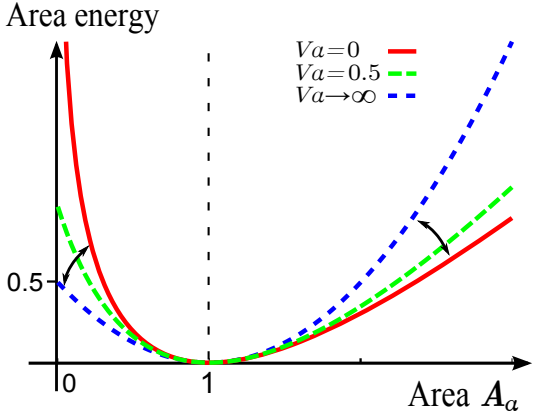}
  \caption{A variation of the area energy of the single cell $\alpha$ with respect to $V_\alpha$. The area is normalized by the least energy area, and the coefficients are given so that the behavior around the minimum coincides with that of the "area constraint" with $\kappa_\alpha^{(2)}=1$. These fix two parameters out of three, and the variation can be given by the value of $V_\alpha$.}
  \label{fig:area_energy}
\end{figure}

Setting $\kappa_\alpha^{(2)}=0$, we substitute Eq. (\ref{def:osmo}) into Eq. (\ref{def:Ea}) and perform the integration of $\Pi_\alpha(S_\alpha)$ over the surface element, where we choose $(1-V_\alpha)$ as its initial point for simplicity. The area part of the energy reduces to
\bea
   E_A &=& \sum_\alpha \left\{ \kappa_\alpha^{(1)} A_\alpha 
    - R_\alpha^{(o)} \log( A_\alpha + V_\alpha ) \right\},
\eea
a term of which is depicted in Fig.\ref{fig:area_energy}. 
Focusing on each contribution from a cell, we easily find its ground state at $A_\alpha= \wt A_\alpha^{(0)}$ where 
\bea
\label{def:target area}
   \wt A_\alpha^{(0)} \equiv -V_\alpha 
     + \frac{R_\alpha^{(o)}}{\kappa_\alpha^{(1)}},
\eea
provided $\kappa_\alpha^{(1)}>0$ and $R_\alpha^{(o)}>0$. For this to be positive, the parameters should satisfy the bound $R_\alpha^{(o)}/\kappa_\alpha^{(1)} > V_\alpha$. We call this $\wt A_\alpha^{(0)}$ the target area that minimizes the area energy of the cell $\alpha$. The buffering area $V_\alpha$ is supposed to be a geometrical constant, while the coupling $\kappa_\alpha^{(1)}$ and $R_\alpha^{(o)}$ may depend on the biochemical environment. Therefore, in principle, we set three parameters, $\{ \kappa_\alpha^{(1)}, R_\alpha^{(o)}, \wt A_\alpha^{(0)} \}$, by fixing two of the three parameters. Expanding each term of $E_A$ around its minimum, it can be approximated by collecting their leading terms as follows:
\bea
  E_A &\simeq& \sum_\alpha \frac{\wt \kappa_\alpha^{(2)}}{2} 
\left( A_\alpha - \wt A_\alpha^{(0)} \right)^2 ,
\nn
 \wt \kappa_\alpha^{(2)} &\equiv&
   \frac{ \left( \kappa_\alpha^{(1)} \right)^2 }{ R_\alpha^{(o)} }, 
\eea
up to a constant. Therefore, even in the absence of the ``area constraint'', the effective energy around the minimum reproduces the form of the ``area constraint''. In fact, one can easily confirm that the above $E_A$ with $\kappa_\alpha^{(2)}=0$ approaches the quadratic form of the constraint as $\kappa_\alpha^{(1)}$ goes to infinity while $\wt \kappa_\alpha^{(2)}$ and $\wt A_\alpha^{(0)}$ are fixed. Consequently, the case with $\kappa_\alpha^{(2)}=0$ and nonzero $\kappa_\alpha^{(1)}$ can be viewed as a generalization of the conventional form, and for this reason, we suppress $\kappa_\alpha^{(2)}$ in the rest of the manuscript.

This generalization adds two novel features to the model: clear physical interpretations and a variety of mechanical responses of the cells. The linear term represents elastic and isotropic tension, while the log term represents isotropic pressure. The ad hoc "area constraint" term would not provide such interpretations, and therefore no mechanical experiments will directly provide the coefficient $\kappa^{(2)}_\alpha$. Secondly, with non-negligible $\Lambda$ and $\Gamma$, or with a sufficient value of $P_{outer}$, the cell areas do not take the minima of their area energies. Therefore, the generalization provides different mechanical responses to the cortical tensions and the external pressure. For example, with a certain set of model parameters and $P_{outer}=0.2$, a simulated tissue with only the ad hoc term collapses, while a tissue with the generalized form may survive. This can be understood in the context of ``cell competition'' \cite{shraiman05} where fast growing cells exert pressures to and may extrude slow growing cells. These simulated tissues will be mentioned in more detail in Section \ref{sec:discussions}.

By some appropriate constant area, the energy can be non-dimensionalized to
\bea
\label{eq:nondimensionalisation}
  E
   &=& \sum_\alpha  \kappa_\alpha \Biggl[
      A_\alpha
    -   R_\alpha^{(o)} \log \left(  A_\alpha +  V_\alpha \right) 
    + \frac{\Gamma_{\alpha}^{(L)} }{2}  L_{\alpha}^2
\nn&&
    + \sum_{\braket{i,j}\in\alpha} \left\{ \frac{ \Lambda_{\alpha \beta}}{2}\,  l_{ij}
    + \frac{ \Gamma_{\alpha\beta}^{(l)} }{4} ( l_{ij})^2 \right\}
   \Biggr] ,
\eea
up to a constant, where the variables and parameters are all non-dimensionalized (see Appendix \ref{app:nondimensionalization}). In the completely homogeneous situation, {\it i.e.}, all the coefficients in Eq. (\ref{def:E}) are common in all cells, the energy can be further reduced to
\bea
\label{def:Ehomo}
  E &=& \kappa \sum_\alpha\Biggl[
    A_\alpha
    - R^{(o)} \log \left( A_\alpha + V_{0} \right) 
    + \frac{\Gamma^{(L)} }{2} L_{\alpha}^2 
\nn&& 
    + \frac{\Lambda}{2}\, L_{\alpha}
    + \sum_{\braket{i,j}\in\alpha} \frac{\Gamma^{(l)} }{4} (l_{ij})^2 
   \Biggr],
\eea
where $\kappa$ denotes the common prefactor and a trivial relation $L_\alpha = \sum_{\braket{i,j}\in\alpha} l_{ij}$ is used. As a reference, we show the single cell energy in the isotropic hexagonal packing with the common edge length $l$: 
\bea
  \varepsilon &=& 
    \left( \frac{3\sqrt3}{2} + 18 \Gamma^{(L)} + \frac{3\Gamma^{(l)} }{2} \right) l^2
    + 3 \Lambda\, l
\nn&&
    - R^{(o)} \log\left(\frac{3\sqrt3}{2} l^2 + V_0 \right).
\eea

\subsubsection{Energy minimization and the Young-Laplace law}
\label{sec:minimisation}

Subject to the potential energy (\ref{def:E}) or its simplified form (\ref{def:Ehomo}) for the homogeneous case, the corresponding forces appear in the system in relation to the dynamical variables. Equating these with the other forces with time derivatives, one reaches the governing equations of motion of the system. In the physics context, 
the equations of motion can be deduced from the corresponding Lagrangian. In particular, the equilibrium states can be found by the minimization of the energy.

In our case, there are two types of independent quantities as candidates for the dynamical variables: the vertex positions and the curvatures of the edges. For the former, the conventional setting for the cellular dynamics assumes some slow dynamics. That is, at the characteristic time scale of the dynamics, the inertial forces are negligible, and only the friction terms appear with the time derivatives of the positions. In this setting, a question arises concerning how the latter variables behave at such a scale. 

In this paper, we suppose that the forces along the edges are well equilibrated in the time scale we examine. 
In other words, we suppose that the force differences at the curved edges with the pressures are relaxed and balanced so swiftly that the curvatures would not appear as dynamical variables in the equations of motion of the system. In short, the curvatures are the fast variables, while the vertex positions are the slow variables. We formulate the slow dynamics of the tissues. In this section, we explain this equilibrium along the line of the minimization of the energy and confirm that the Young-Laplace law arises as its natural consequence. For the precise and detailed derivation of the formulas we use, see Appendix \ref{app:pdA}.

Let us first examine the variation of the energy, $\delta E$. The minimization of the energy can be accomplished by the vanishing virtual work done by arbitrary changes in the variables, that is, $\delta E = 0$. In addition, the potential part of the Lagrangian is sufficient for our purpose because we only need to clarify the force equilibrium at the edges. Because of the decomposition (\ref{def:EaEl}) of the energy, the variation of the energy can be given by the variations of $A_\alpha$ and $l_I$:
\bea
  \delta E &=& -\sum_\alpha p_\alpha \delta A_\alpha + \sum_I T_I \delta l_I , 
\eea
where $p_\alpha \equiv - \frac{\pa E}{\pa A_\alpha}$ and $T_I \equiv \frac{\pa E}{\pa l_I}$ give the formal definitions of the pressures and the tensions, respectively. In the above, the numbering of the edges by a single capital letter is introduced, while the orientation of the edges is implied by the convention of the sign of the curvatures. Note that the variation of the energy is given primarily by $\delta \x_i$ and $\delta R_I$, and $\delta A_\alpha$ and $\delta l_I$ are the functions of them. The sum over the cell numbering $\alpha$ can be distributed to the one over the edge numbering by the resummation formula (\ref{formula:resummation}) shown in Appendix \ref{app:pdA}:
\bea
  \delta E 
   &=& \sum_I \left[ T_I \delta l_I - \left( p_{\alpha_I} - p_{\beta_I} \right) \delta A_{\alpha_I,I} \right],
\eea
where $\alpha_I$ and $\beta_I$ are the cells to the left and right of the edge $I$, respectively. $\delta A_{\alpha_I,I}$ denotes the $I$-th edge fraction of the variation of the cell area. One can further reduce the edge fraction of $\delta A_{\alpha_I,I}$ to the following three terms (see Appendix \ref{app:pdA}).
\bea
  \delta A_{\alpha_I,I}
   &=&  R_I \delta l_I
    + R_I \sin\theta_I \delta l_I^{\perp} 
    - R_I \cos\theta_I \delta l_I^{(0)} ,
\eea
where $\delta l_I^{(0)}$ is the change in the distance between two ends of the edge $I$ and $\delta l_I^{\perp}$ is pointing outward to the cell $\alpha_I$, perpendicular to $\delta l_I^{(0)}$. Hence, the work done by the change in the vertex positions and the curvatures is given by
\bea
\label{eq:deltaE}
  \delta E 
  &=& \sum_I \Bigl[ 
    \left\{ T_I - (p_{\alpha_I} - p_{\beta_I}) R_I \right\} \delta l_I
\nn&&
    - (p_{\alpha_I} - p_{\beta_I}) R_I 
      \left( \sin\theta_I \delta l_I^{\perp} - \cos\theta_I \delta l_I^{(0)} \right)
  \Bigr].
\eea

The variations of the curvatures are only contained in the first term, and thus the minimization of the energy, $\delta E=0$, in terms of the curvatures requires the vanishing coefficients of the first term and yields the following Young-Laplace law for each edge:
\bea
\label{def:YLlaw}
   \frac{T_I}{R_I} &=& p_{\alpha_I} - p_{\beta_I}.
\eea
As in the derivation, the tension may take a negative value. However, by considering the buckling effect in the absence of the bending rigidity, we exclude such a situation from our current consideration, as will be briefly mentioned again in the determination of the curvatures. In the force equilibrium along the edges, the variation of the energy can be much simplified by substituting the Young-Laplace law into Eq. (\ref{eq:deltaE}).
\bea
\label{eq:deltaE-YL}
  \delta E &=& - \sum_I T_I \left( \sin\theta_I \delta l_I^{\perp} - \cos\theta_I \delta l_I^{(0)} \right).
\eea
This fact will be reflected in the derivation of the EOMs in the following section.

Although the variation of the energy takes the derived form in general, there arises a different form of $\delta E$ in the case with the saturated curvatures. In such a case, the saturation behaves as the constraint that prevents the curvature from canceling the pressure difference along the edge. In terms of the variation of the energy, this can be viewed as an excess energy over the Young-Laplace law. When the $I$-th edge is saturated, $|\theta_I|=\frac{\pi}2$, and consequently, $0 = \delta (l_I^{(0)}/(2R_I))$. This yields the following relationship at the saturation:
\bea
  |R_I| = \frac{l_I^{(0)}}{2} , \quad 
  R_I \delta l_I^{(0)} = l_I^{(0)} \delta R_I.
\eea
Accordingly, the variation of the edge length $l_I$ takes a different form for saturated $|\theta_I|=\frac{\pi}{2}$:
\bea
 \delta l_I &=& 2\, \delta R_I \theta_I 
  = \frac{2 R_I \theta_I}{l_I^{(0)}} \delta l_I^{(0)}
  = \frac{\pi}{2} \delta l_I^{(0)}.
\eea
Therefore, the $I$-th part of the variation of the energy becomes
\bea
\label{eq:deltaE-YL-saturated}
  \delta E_I &=&
   - T_I \sgn(\theta_I) \delta l_I^{\perp}
\nn&&
   - \left\{ (p_{\alpha_I} - p_{\beta_I}) - \frac{T_I}{R_I}\right\} 
      R_I  
      \left( \frac{\pi}{2} \delta l_I^{(0)} + \sgn(\theta_I)\delta l_I^{\perp} \right).
\nn
\eea
The first term takes the same form as that for the unsaturated edges, whereas the second term shows the excess energy and its corresponding force. The excess force pushes the relevant vertices toward the direction in which the edge becomes isotropically swollen. As such, in spite of the artifact of the saturation bounds for the curvatures, the resulting excess force arises in a mechanically sound manner and is energetically consistent with the possible interpretations mentioned in Section \ref{sec:energy}.

\subsubsection{The equations of motion}


The rather conventional form of the EOMs of the vertex positions as the governing equations of the dynamics are defined as follows. Equating the friction terms and the potential forces leads to the following:
\bea
\label{def:EOMs}
  \eta_i \, \frac{d \left(\x_i -\Vec{X}_i\right) }{dt} = - \frac{\pa E (\{\x_i\})}{\pa \x_i}.
\eea
The left side of the equation is the friction, while the right side is the potential force. The coefficient $\eta_i$ is the friction coefficient for the vertex $i$, and the variable $\Vec{X}_i$ is introduced as a referential position to the vertex $i$. The noise term is suppressed for convenience, and the inertial terms are neglected, as stated previously. Precisely speaking, the friction term should be of the relative positions or of the relative velocities to the surface or the environment to which the friction is actually occurring. In a biological context, the friction term is dependent on the conditions under which the experiments are conducted. For example, if the epithelial cells are attached to some firm substrate, then $\Vec{X}_i$ should be given mostly by the position of the substrate. Under such a condition, they all take nearly the same constant and are therefore negligible as the zero vector. By contrast, if the cells are freely movable to in the surrounding medium, then $\Vec{X}_i$ is supposed to be some effective position of the surroundings, which may be given by an averaged quantity of the neighboring cell positions in a sophisticated way. For simplicity, in the following we consider the former case, $\Vec{X}_i\simeq 0$, with a common friction coefficient $\eta_i = \eta$. Note that, once the system reaches equilibrium, we do not need to consider the difference due to the vanishing frictions, although the vanishing frictions do not necessarily compensate the equivalent equilibrium states in both cases.

The EOMs (\ref{def:EOMs}) appear formally identical for other existing vertex models \cite{oster90,honda01,honda04,nagai88,okuda13}. However, the explicit form of the above is quite distinct in our case, mainly due to the presence of the curvatures. In the other models, edges behave mechanically as elastic but rigid objects, such that the internal pressure of a cell directly pushes the vertex positions of the cell. In our model, in the absence of the rigidity, the pressure pushes the edge and bends it until the pressure is balanced with the tension along the edge. Thus, the pressure exerts its force onto the vertices only through the tension. This can be clearly seen in Eq. (\ref{eq:deltaE-YL}), and the potential forces can be accounted for solely by the tensions. In fact, the potential forces are derived by differentiating the energy in terms of the vertex positions:
\bea
\label{def:pot force}
  - \frac{\pa E(\left\{ \x_i \right\})}{\pa \x_i}
 &=& \sum_I T_I \left( \sin\theta_I \, \frac{\pa l_I^{\perp}}{\pa \x_i}
    - \cos\theta_I \, \frac{\pa l_I^{(0)}}{\pa \x_i} \right). \phantom{m}
\eea
Because only three edges connect to a vertex, the summation is over the three edges, each of which amounts to a tension toward the direction the edge is connected to. Substituting the potential forces (\ref{def:pot force}) into the EOMs (\ref{def:EOMs}) with $\eta_i=\eta$ and $\Vec{X}_i=0$, it leads to
\bea
\label{eq:EOMs}
  \eta\, \frac{d \x_i}{dt} &=& \sum_I T_I \left( \sin\theta_I \, \frac{\pa l_I^{\perp}}{\pa \x_i}
    - \cos\theta_I \, \frac{\pa l_I^{(0)}}{\pa \x_i} \right).
\eea
The explicit form of the tension can be written by
\bea
\label{def:T} \label{eq:T}
  T_I = \ol{\Gamma}_{\alpha\beta} l_I + \ol \Lambda_{\alpha\beta} ,
\eea
where $\alpha$ and $\beta$ represent the cells to the left and the right of the edge $I$, respectively, and
\bea
\label{def:Tcoeff}
  \ol \Gamma_{\alpha\beta} &\equiv& 
  \Gamma_{\alpha\beta}^{(l)} + \left( \Gamma_{\alpha}^{(L)} + \Gamma_{\beta}^{(L)} \right) , 
\nn
  \ol \Lambda_{\alpha\beta} &\equiv&
    \Lambda_{\alpha\beta} + \left\{
    \Gamma_{\alpha}^{(L)} \left( L_\alpha - l_I \right)
    + \Gamma_{\beta}^{(L)}  \left( L_\beta - l_I \right)
    \right\} .
\eea
The second term of $\ol \Lambda_{\alpha\beta}$ is obtained from the decomposition of the perimeter term to linear and constant terms in $l_I$. That is, $\left(L_\alpha-l_I\right)$ is an $l_I$-independent combination, as is the factor $\ol \Lambda_{\alpha\beta}$. Because the length of the edge satisfies the inequality, $l_I \geq l_I^{(0)}$, the assumption of the non-negative tension can be read as $\ol \Gamma_{\alpha\beta} l_I^{(0)} \geq -\ol \Lambda_{\alpha\beta}$.

In the case with the saturated curvatures, we follow the same approach. Substituting the coefficients in Eq. (\ref{eq:deltaE-YL-saturated}) into the potential force, the EOMs become
\bea
\label{def:EOMsaturated}
  \eta\, \frac{d \x_i}{dt} &\!=\!& \sum_I T_I \left( \sin\theta_I \, \frac{\pa l_I^{\perp}}{\pa \x_i}
    - \cos\theta_I \, \frac{\pa l_I^{(0)}}{\pa \x_i} \right)
\nn&&\!\!
   + \sum_J \Biggl[ T_J \sgn(\theta_J) \delta l_J^{\perp}
\nn&&\!\!
   + \left\{ (p_{\alpha_J} \!\!-\! p_{\beta_J}) \!-\! \frac{T_J}{R_J}\right\} \!
      R_J \!
      \left( \frac{\pi}{2} \delta l_J^{(0)} \!+\! \sgn(\theta_J) \delta l_J^{\perp} \right)
    \Biggr] ,
\nn
\eea
where $I$ represents the unsaturated edges and $J$ represents the saturated edges:
$R_J=\sgn(\theta_J) l_J^{(0)}/2$.

\subsubsection{Determination of curvatures by the Young-Laplace law}
\label{sec:determination}

So far, we have seen that the minimization of the energy (\ref{def:E}) in terms of the curvatures leads to the Young-Laplace law (\ref{def:YLlaw}) for each edge. It follows that the equations of motion take the form (\ref{eq:EOMs}), with the vertex positions as the only remaining dynamical variables. Thus, in our current formalism, the curvatures are determined at any time $t$ by the law for the given set of the model parameters and the positions $\{\x_i(t)\}$. On the other hand, the values of the curvatures are required for the explicit form of the EOMs because it is clear in Eq. (\ref{eq:T}) that the tensions in the EOMs depend on the values of the curvatures. Therefore, to complete the basic construction of the dynamics, we must clearly describe how the law determines the curvatures for the given set of the parameters and the vertex positions.

First, we must confess that there is no known method, in general, to obtain the analytic forms of the curvatures as the exact solutions of the Young-Laplace equations (\ref{def:YLlaw}). The curvatures or radii of the curvatures appear nonlinearly in the definitions (\ref{def:l}) of the lengths and areas and, therefore, in the tensions and pressures as well. Accordingly, the Young-Laplace equations emerge as a set of non-linear equations for the curvatures whose number equals the number of edges. For $N_v$ vertices, $N_e$ edges, and $N_c$ cells, Euler's formula states $N_v -N_e +N_c =1$, and vertices with degree three mean $3N_v=2N_e$. Therefore, $N_e$ amounts to
\bea
  N_e = 3 ( N_c - 1 ).
\eea 
The number of equations is approximately threefold that of the number of cells. From this practical circumstance and the aforementioned nonlinearity, we infer that there is no means of finding a solution in a rigorous way or even assuring the existence of a solution.

Instead of searching for an exact solution, we intend to find a numerical solution of the Young-Laplace equations, assuming that a solution exists. Our tactics is as follows. In the equational form of the law (\ref{def:YLlaw}), we unite both sides of the equations into one form as a set of the functions of the curvatures: $\{G_I\}$, for $I$ being the edge number,
\bea
\label{def:YLfunc}
  G_I &\equiv& \frac{T_I}{R_I} - \left( p_{\alpha_I} - p_{\beta_I} \right).
\eea
The suffixes $\alpha_I$ and $\beta_I$ represent the cells to the left and right of  the edge $I$. We will henceforth refer to these functions as the Young-Laplace functions. Thus, the law is rewritten as $G_I = 0$, and the problem of finding the solution is redefined by that of finding the simultaneous zeroes of $\{G_I\}$. Investigating the analytic behaviors of the functions with respect to the curvatures, we propose an optimization algorithm of the squared sum of the functions to find the numerical solution. The algorithm constitutes the method of steepest descent and a simple collection of the bisection methods, the latter of which is only occasionally used to complement the former. Note that some modification to the definitions of the functions will be introduced due to the presence of the bounds of the curvatures. Note also that the algorithm is strongly supported by the facts that, in many cases, the solutions are all isolated and there is only one solution in the analytic domain of the functions. The proofs and the limitations on the solution are provided in Appendix \ref{app:proofs}.

Let us show the explicit form of the Young-Laplace functions $\{G_I\}$. The internal pressure $p_\alpha$ of the cell $\alpha$ is defined by the minus gradient of the energy in terms of the area $A_\alpha$, which corresponds to the volume in three dimensions. It consists of the elastic part $p_\alpha^{(el)}$ and the osmotic part $\Pi_\alpha$:
\bea
\label{def:p-decomposition}
  p_\alpha(A_\alpha) \equiv - \frac{\pa E}{\pa A_\alpha}
   = p_\alpha^{(el)}(A_\alpha) + \Pi_\alpha (A_\alpha) ,
\eea
where
\bea
\label{def:p-components}
  p_\alpha^{(el)} 
   &=& -\kappa_\alpha^{(1)}
       - \kappa_\alpha^{(2)} \left( A_\alpha - A_\alpha^{(0)} \right) ,
\nn
  \Pi_\alpha 
   &=& \frac{ R_\alpha^{(o)} }{ A_\alpha + V_\alpha }. 
\eea
The coefficients are not yet non-dimensionalized to show $\kappa_\alpha^{(1)}$ and $\kappa_\alpha^{(2)}$ together. Inserting the expressions (\ref{def:T},\ref{def:p-decomposition},\ref{def:p-components}) into Eq. (\ref{def:YLfunc}), the Young-Laplace functions turn out to be:
\bea
\label{eq:Ggeneral}
  G_I 
  &=& \frac{\ol \Gamma_{\alpha\beta} l_I + \ol \Lambda_{\alpha\beta}}{R_I}  - P_{const}
\nn&&
  + \left( \kappa_\alpha^{(2)} A_\alpha - \kappa_\beta^{(2)} A_\beta \right) 
    - \left( \frac{ R_\alpha^{(o)} }{ A_\alpha + V_\alpha }
       - \frac{ R_\beta^{(o)} }{ A_\beta + V_{\beta} } \right)
\nn
\eea
with the constant term $P_{const} \equiv  \kappa_\alpha^{(2)} A_\alpha^{(0)} - \kappa_\beta^{(2)} A_\beta^{(0)} - \kappa_\alpha^{(1)} + \kappa_\beta^{(1)}$ for edge $I$ in the bulk. In the homogeneous case, $P_{const}$ and the dependence on $\kappa_\alpha^{(1)}$ vanish except for the boundary edges. For the edges at the boundary of the tissue, one of the pressures should be replaced by $P_{outer}$ in Eqs. (\ref{def:YLfunc},\ref{eq:Ggeneral}), which is constant with respect to the curvatures. Writing the functions in vectorial form, the law takes a rather simple form:
\bea
\label{def:vYL}
  \Vec{G} &=& \Vec{0}.
\eea
This can be regarded as the implicit equations in the domain of the function. Because the number of equations is equal to that of the edges, the solution is zero-dimensional, and one may expect it is a point in the domain. 

Next, we redefine the variables and give a brief description of the domain of the function for our analysis. For the detail of the domain, see Appendix \ref{app:domain}. The curvatures are to be determined at each time by the law. Thus, $\Vec{G}$ is the vectorial function of $\{R_I\}$. For later convenience, we introduce the vectorial variable $\Vec{\rho}$ of the normalized curvatures, for which each edge component is
\bea
\label{def:normalised-variable}
  \rho_I \equiv \frac{l_I^{(0)}}{2 R_I}.
\eea
The range of the value is normalized to $-1\leq \rho_I \leq 1$ due to the saturation bounds of the curvatures. Denoting the interval by $D\equiv [-1,1]$, the domain of the function becomes, at most, the $N_e$-dimensional hypercube $D^{N_e}$ for $N_e$ cell boundaries. When the osmotic pressure is divergent or the cell area has a zero for $\rho_I \in D$, the interval should be decreased to $\wt D \subseteq D$. Therefore, $\Vec{G}$ is a function of $\Vec{\rho}$ in the domain $\wt D^{N_e}$.

\begin{figure}[h]
  \centering
  \includegraphics[width=6.0cm]{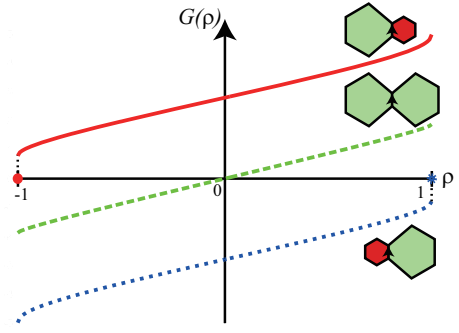}
  \caption{A schematic picture of the modified Young-Laplace function $G(\rho)$ of the central edge with its normalized curvature $\rho$ in three typical cases. The orientations of the central edges are shown in the associated diagrams. 
The two saturated cases are modified at $\rho=-1,1$, respectively.
}
  \label{fig:Gmodified}
\end{figure}

As noted previously, the definition of the function $\Vec{G}$ must be modified due to the presence of the bounds of the curvatures. When no zero of $G_I$ appears in the interval $\wt D$ for $\rho_I$, no solution seems to exist. However, this is an artifact because the bounds of the curvatures block $\rho_I$ from running to the value satisfying $G_I=0$. Therefore, the function must be modified according to a physical interpretation of no zero. Namely, we force $G_I$ to be zero at the end of $\wt D$ for $\rho_I$ in the cases in which $G_I$ takes only negative, or positive, values. 
\bea
\label{def:Gmodification}
  G_I(\rho_I = \max(\rho_I)) &=& 0 \quad {\rm if~} G_I(\rho_I < \max(\rho_I) ) < 0 ,
\nn
  G_I(\rho_I = \min(\rho_I)) &=& 0 \quad {\rm if~} G_I(\rho_I > \min(\rho_I)) < 0 .
\eea
When $G_I$ is always negative for $\rho_I \in \wt D$ for given $\{\rho_{J\neq I}\}$, it physically means that the pressure difference is positively too large to be compensated by the tension. Thus, the cell boundary becomes maximally swollen, and $\rho_I$ takes its upper limit: $\rho_I=\max(\rho_I)$. For this to be consistent with the equation, $G_I$ must satisfy $G_I(\rho_I = \max(\rho_I)) = 0$, regardless of its continuity. Similarly, $G_I(\rho_I = \min(\rho_I)) = 0$ in the positive $G_I$ case. When $G_I$ takes both negative and positive values in $\wt D$ for $\rho_I$, there is always a solution of $G_I=0$ inside the interval, thanks to Bolzano's theorem\footnote{Bolzano's theorem is a special case of the intermediate value theorem.}. By this prescription, one can find a value of $\rho_I$ that meets $G_I=0$ for any given values of $\{\rho_{J\neq I}\}$. In return, the function $\Vec{G}$ becomes analytic only in $\wt D^{N_e}$, except at the above special boundary defined by the modification. The above three cases are schematically depicted in Fig. \ref{fig:Gmodified}. 
In what follows, we refer to the function $\Vec{G}$ as the modified function unless otherwise stated.

Now, we are ready to analyze the Young-Laplace function $\Vec{G}$ and justify the use of our optimization on the squared sum of the functions, {\it i.e.}, $\left| \Vec{G}\right|^2$. 
First, one can prove that $G_I$ is a monotonically increasing function of $\rho_I$ in the analytic domain of $\wt D^{N_e}$, and this increasing nature persists in the entire domain. In addition, we may expect that $G_I$ is either a constant or a monotonic function of other variables in many cases (see Appendix \ref{app:monotonicity} for the proof and the limitations). Accordingly, a simple bisection method can easily find the numerical value of $\rho_I$ such that $G_I(\rho_I)=0$ for fixed values of $\{\rho_{J\neq I}\}$, although it is not suitable for finding the solution of $\Vec{G}(\Vec{\rho})=\Vec{0}$. Most importantly, $G_I^2$ is a downward-convex function of $\rho_I$ and can be expected to be either a constant or a downward-convex function of $\rho_{J \neq I}$. For this reason, we may expect that the squared sum of the functions, {\it i.e.}, $\left| \Vec{G} \right|^2$, is a downward-convex function of $\Vec{\rho}$.

On the basis of the above, we propose the optimization algorithm as follows. First, we set the following evaluation function $U$ to be minimized:
\bea
  U \equiv \left| \Vec{G} \right|^2.
\eea
It is obvious that $U\geq 0$ and the global minimum $U=0$ give the solution of Eq. (\ref{def:vYL}). To find the roots, because the vertex positions are fixed, the distances between two ends of the edges $\{l_I^{(0)}\}$ and the polygonal areas $\{ A_\alpha^{(poly)} \}$ are given. Next, we use the value of $\Vec{\rho}$ at the time $t-\delta t$ for infinitesimally small $\delta t>0$ as the initial point for finding the minimum of $U$ at time $t$. Because we are dealing with the slow dynamics, we can also expect the value of $\Vec{\rho}$ at $t$ would not change substantially from the values at $t-\delta t$. If we cannot use $\Vec{\rho}$ at $t-\delta t$, for example, at the initial time, it would be appropriate to use the values calculated by the simple collection of the bisection methods mentioned above. Finally, by the method of steepest descent, we search the numerical solution, shifting $\Vec{\rho}$ toward $\Vec{v}$:
\bea
\label{def:SDM}
  \Vec{v} &=& - \nabla \left| \Vec{G} \right|^2 = - 2\, J^T\, \Vec{G},
\eea
where $\nabla$ denotes the gradient in the $N_e$-dimensional functional domain: $\nabla\equiv \left( \frac{\pa}{\pa\rho_1}, \ldots, \frac{\pa}{\pa\rho_{N_e}} \right)^T$. $J$ is the Jacobian matrix as a matrix-valued function of $\Vec{\rho}$ such that $J_{IJ} \equiv \frac{\pa G_I(\Vec{\rho})}{\pa \rho_J}$. There is an important note and a modification of the above expression because we have modified the functions in Eq. (\ref{def:Gmodification}). That is, when the edge $I$ is saturated, the function $G_I$ might be discontinuously modified so that we cannot define the derivatives around the point. In such a case, the saturated $\rho_I$ should not be shifted by the method so that $v_I$ be zero. In addition, the discontinuously saturated edge should also be discontinuously saturated when $\rho_{J\ne I}$ is infinitesimally changed because the original form of $G_I$ is analytic. Mathematically, this means that $\frac{\pa G_I}{\pa \rho_{J\neq I}}=0$. These lead to the corresponding modification to the above as
\bea
  \Vec{v}_I &=& 0,
\nn
  \frac{\pa G_I}{\pa \rho_{J\neq I}} &=& 0,
\eea
when the $I$-th edge is discontinuously saturated as prescribed by Eq. (\ref{def:Gmodification}). In numerical calculations, these can easily be implemented by eliminating the relevant terms in $\Vec{v}$ by $v_I=0$ and $G_I=0$.

Finally, we would like to remark that there are a variety of numerical methods to solve a set of non-linear equations. Among them, the Newton-Raphson method, a.k.a. Newton's method, prevails due to its ease of use and the speed of convergence. One can utilize this instead of the method of steepest descent. However, Newton's method runs away from the true solution when the second derivative of the function changes its signature. In our case, the function $\Vec{G}$ is highly nontrivial, and the derivatives of the function actually include some hypergeometric functions that may change the signs of the derivatives. In fact, we have confirmed that Newton's method does not always reduce $U$. Another candidate is the conjugate gradient method, which uses the previous search. This method is quite adequate when the method of steepest descent hardly pushes $U$ toward zero but may require far more computational effort than the method of steepest descent. 

\subsubsection{Topological changes}
\label{sec:topological changes}

Epithelial tissues, or a collection of dry foams, undergo various types of topological changes during their dynamical processes, including cell rearrangements, cell apoptosis or extrusions, cell divisions (or coalescence in foams), and transformations of a vertex with degree $n$ to one with different degree. These changes are of great importance for the mechanical nature of the tissues, {\it i.e.}, their viscoelastic and plastic nature, particularly under deformations or in morphogenesis.

In this section, we describe the topological changes and how to implement them in our model to finalize our construction. To be precise, we deal with the processes regarding the cell rearrangements, cell disappearances (or apoptoses), and cell divisions. As stated previously, the transformations of vertices with different degrees are suppressed by our current setting in which the degrees of the vertices are all three. 

\begin{figure}[h]
  \centering
  \includegraphics[width=8.0cm]{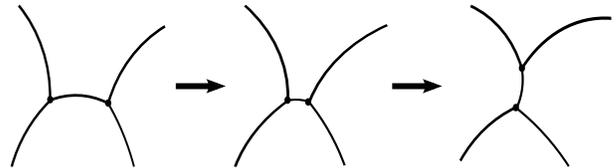}
  \caption{An illustration of the T1 topological change.}
  \label{fig:T1}
\end{figure}

\begin{figure}[h]
  \centering
  \includegraphics[height=2.5cm]{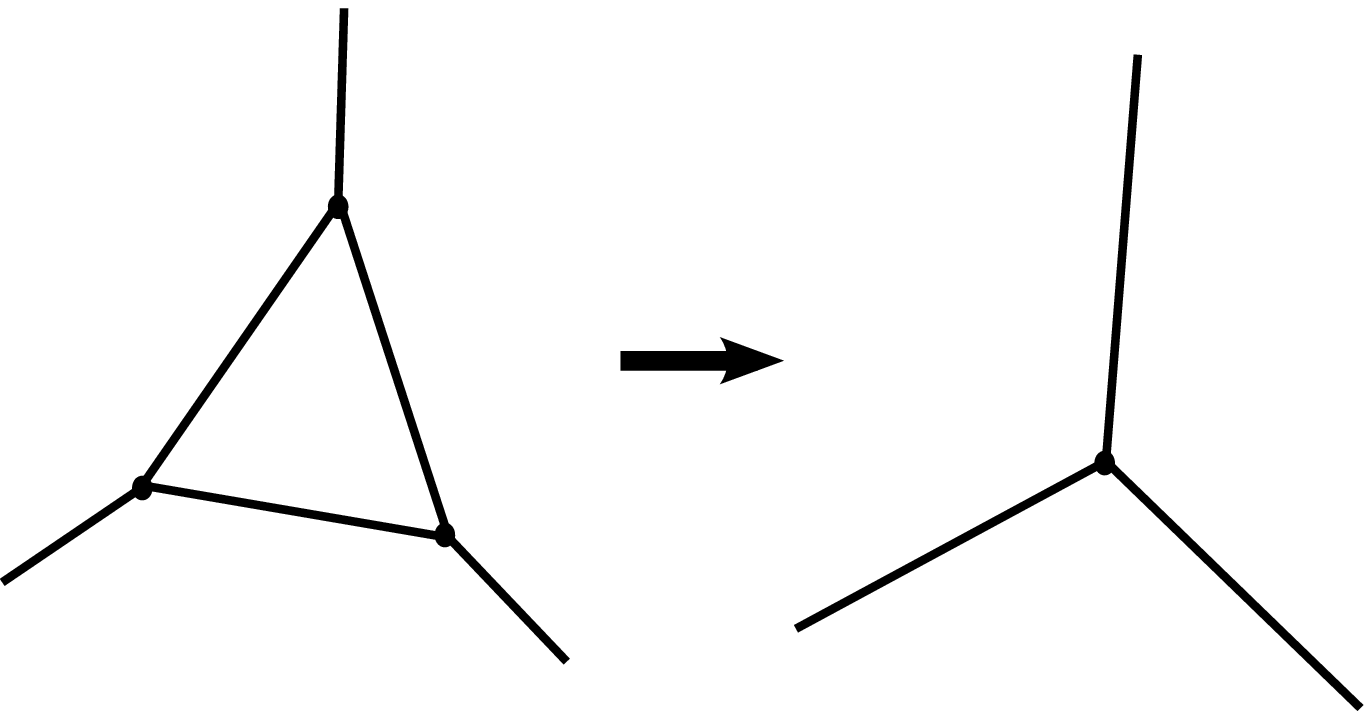}
  \caption{An illustration of the T2 topological change.}
  \label{fig:T2}
\end{figure}

In two-dimensional vertex models, the cell rearrangements and the cell disappearances can be described by means of the so-called T1 and T2 topological changes \cite{weaire84}. Namely, when a cell boundary shrinks, recombination of the vertices may occur, and the cells may exchange their neighbors around the cell boundary. This topological change is called T1 (Fig. \ref{fig:T1}) and is referred to as the switching process in \cite{honda01}. The T2 change represents the disappearance of a triangular cell from the defined geometry (Fig. \ref{fig:T2}). 

\begin{figure}[h]
  \centering
  \includegraphics[height=2.5cm]{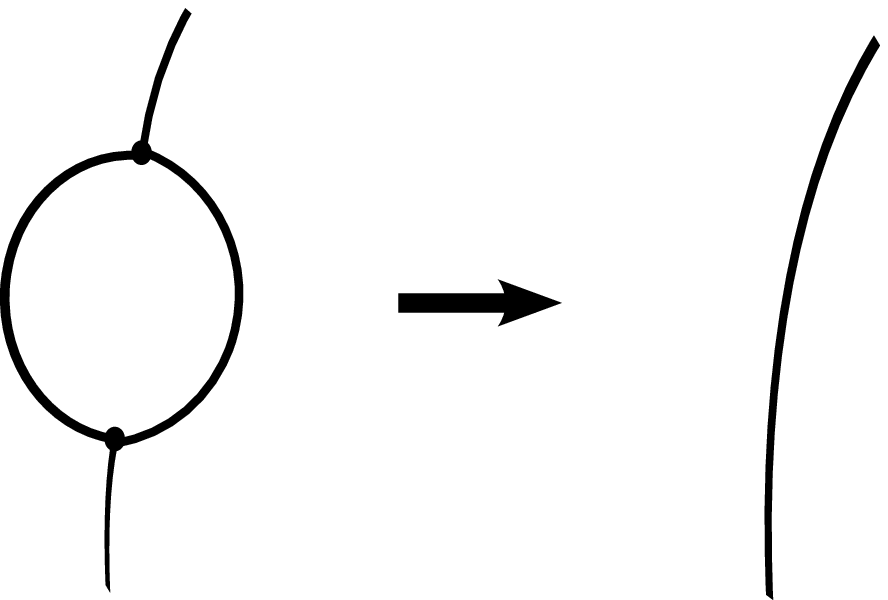}
  \caption{An illustration of the T2' topological change.}
  \label{fig:T2'}
\end{figure}

In the BVD, the T1 topological change involves curved edges. Because T1 occurs at a fairly short edge, we first keep the curvatures of the other edges intact and recalculate them after the recombination. The T2 change notably differentiates our model from other existing vertex models. A T2 process lets a cell disappear from the geometry such that the cell either dies or is extruded from. If we scrutinize the disappearance process in time, T1 precedes T2 as a reduction of the vertices of the cell up to the minimum number of edges, and T2 finalizes the disappearance as in Fig. \ref{fig:T2}. In the polygonal models, triangle has the minimum number of edges, so it is sufficient to define the T2 process by disappearing triangular cells. However, in our model, two-vertex cell \footnote{There is a term ``biangle'' for a shape with two vertices only, but it is by definition a polygon on a spherical geometry. Therefore, to avoid confusion, we use the term ``two-vertex cell'' rather than biangle.} has the minimum number of vertices. Therefore, our T2 topological process, or T2' process, should be given by the disappearance of a two-vertex cell as in Fig. \ref{fig:T2'}. 

There are a variety of ways to implement these processes in simulation. The simplest method is to introduce certain thresholds for the processes below which the topological changes are forced to occur. The other method is to let the changes occur if they are energetically favored. In practice, some thresholds below which the energies of the relevant topologies should be compared would also be needed for the latter method. In the next section, in which we demonstrate the model, we employ the former method for simplicity. We introduce the threshold of the distance $\Delta_1$ of two connected vertices, below which the T1 process is forced to occur. In addition, another threshold length $\Delta_2$ is introduced such that if the distance between two vertices of a two-vertex cell decreases below the threshold, the cell disappears. Note that, in the T1 process, a cell boundary is replaced by a newly created boundary with the new curvature variable that is to be fixed immediately. The easiest method is to apply the bisection method to the new curvature because the process represents infinitesimal changes in a pair of vertex positions, and the surrounding pressures and tensions will not change significantly. However, we fully solve the curvatures numerically in the following section.

In biological experiments, cell division is closely related to tissue growth. Divisions are accompanied by cell growth (in area in two dimensions), and the division planes might be strongly correlated with themselves or the tissue environment. Thus, the mechanisms of the divisions should be contemplated with this information. The division itself can be realized simply by adding an edge and a pair of vertices to divide the cell, which changes the variable set of the energy function and, consequently, the energy itself. Therefore even if the division is due only to biological processes, the process is immediately followed by the dynamical relaxation process of the changed energy, which is usually uplifted by the division. Therefore, for example, the division plane will be immediately slanted according to the energy landscape. 

As explained above, there could be a number of different ways to implement the cell division. For example, all cells might grow subject to constantly increasing target areas, then divide themselves when they exceed some threshold area. Alternatively, a cell might divide if it is elongated enough along its longest axis so that the tissue retains some uniformity in cell size and shape, up to some extent. In the next section, we adopt another method by introducing cell cycles for each cell with some randomness. The cell cycle prescribes the schedule of the cell life, including cell growth in area and cell division. For our model, an increase in the tension of the division plane is also introduced immediately after the emergence of the division plane, to model the mitotic cell rounding of the eukaryotic cells \cite{stewart11}.

\subsection{Examples by simulation}
\label{sec:examples}
\subsubsection{Settings for the simulation}

\begin{figure}[h]
  \centering
  \includegraphics[width=8.5cm]{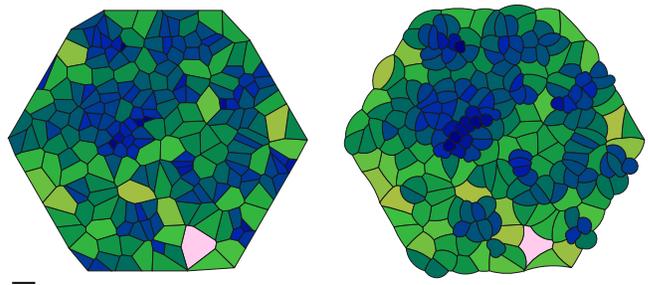}
  \caption{The initial configurations of the vertices and the curvatures used for the examples. The initial positions of the vertices are given by the Voronoi diagram (LEFT). The initial configuration of the curvatures (RIGHT) is determined by the simple collection of the bisection methods and the method of steepest descent for the parameters (\ref{eq:params}). The bar at the bottom represents the unit length. In epithelial tissues, the unit length is approximately $O(10 \mu{\rm m})$. The colors (tones) represent relative values of cell areas within an image: the darker the color is, the smaller the cell area is.}
  \label{fig:init}
\end{figure}

In this section, we present four examples of the proposed model obtained by computer simulation. The geometry is commonly set to a two-dimensional flat surface with a boundary, which is topologically equivalent to a disk. No particular boundary condition is imposed onto the boundary, {\it i.e.}, it is the free boundary condition with the constant pressure of the outer region of the tissue: $P_{outer} = 0.2$. No other external force is applied. This means that there are no specific conditions for vertex positions and velocities at the boundary of the tissue, and they obey the same EOMs as in the bulk. The initial numbers of cells, vertices, and edges are set as
\bea
\label{eq:nums}
  N_c = 217, \quad N_v = 432, \quad N_e = 648.
\eea 
The initial configuration of the vertex positions is generated from a Voronoi diagram obtained from randomly distributed Voronoi centers in a regular hexagon (Fig. \ref{fig:init}) \footnote{
We have tested square and hexagonal patches for the initial configurations, and we choose hexagonal ones because they are more suitable than the others for counting non-peripheral cells.
}. For the generation of the graph with degree three, the corners of the hexagon are removed, and their adjacent edges are united. Then, the initial value of $\Vec{\rho}$ in $\wt D^{N_e}$ is first generated by a simple collection of the bisection methods for individual edges and is reduced by the method of steepest descent as in Fig. \ref{fig:init} before the dynamical step starts. As a matter of course, we must employ the numerical method with some specific criteria and the model parameters; these criteria and parameters will be described below. The average area is set to approximately $0.5$ of the unit target area, and thus we intend to perform simulations of slightly expanding tissues.

In the simplest case with the homogeneity, the energy function is simplified to Eq. (\ref{def:Ehomo}), and the number of fundamental parameters is reduced to thirteen: nine parameters for the energy function, two for the EOMs (\ref{def:EOMs}), and two for the topological changes. Basically, we assume this homogeneity and set the parameters as
\bea
\label{eq:params}
   \Lambda_{\alpha\beta} &=& 0.12, 
\quad 
   \Gamma_{\alpha\beta}^{(l)} \,=\, 0.03,
\quad 
   \Gamma_{\alpha\beta}^{(L)} \,=\, 0.03,
\nn
   \kappa_\alpha^{(1)} &=& 1,
\quad 
   \kappa_\alpha^{(2)} \,=\, 0,
\quad 
   A_\alpha^{(0)} \,=\, 0,
\nn
   R_\alpha^{(o)} &=& 1,
\quad 
   V_\alpha \,=\, 0,
\quad 
   P_{outer} \,=\, 0.2,
\eea
unless otherwise stated. We also set a special value of $\Gamma_\alpha^{l}=0.06$ for the outermost edges to make the tissue tight. The model is already non-dimensionalized with the above and so is the natural unit of length as unity. The average length of the edges is roughly of the same order. To perform the simulation, the EOMs (\ref{def:EOMs}) are discretized, and we use Euler's method as below.
\bea
\label{def:EOMdiscrete}
  \x_i(t+\Delta t) &=& \x_i(t) + \Delta \x_i(t),
\nn
  \Delta \x_i(t) &=& - \frac{1}{\eta} \left( \frac{\pa E(t)}{\pa \x_i} \right) \times \Delta t,
\eea
where $\Vec{X}_i = 0$ and $\eta_i = \eta$ is homogeneously given, as stated above. $\Delta t$ and $\Delta \x_i$ denote the time step and the discretized changes in the vertex positions. In the examples, we set 
$$\eta = 1 \quad {\rm and} \quad \Delta t = 0.01,$$
and we occasionally reduce $\Delta t = 0.001$ if the estimated value of $\left| \Delta\x_i \right|$ is too large. The threshold for the T1 process is also introduced by:
\bea
  \Delta_1 = 0.01,
\eea
whereas $\Delta_2$ for the T2 is omitted. Because the buffering area for the osmotic pressure is suppressed by $V_\alpha = 0$, the osmotic pressure diverges as the cell area shrinks. Accordingly, by definition, the T2 process never occurs.

To compute the potential force in Eq. (\ref{def:EOMdiscrete}) and generate the initial configuration of the curvatures, the curvatures at time $t$ are calculated numerically by the method of steepest descent. Although the method assures that the resulting $U$ approaches a local minimum, which is supposed to be the global minimum in many cases, the speed of convergence is expected to be rather slow. Therefore, we set two criteria to end the calculation: the maximum number of iterations and the satisfactory precision of $U$. When either of these criteria are satisfied, the iterative calculation is aborted, and the final value of $\Vec{\rho}$ is taken as an approximate solution. The maximum number of iterations is set to $10^4$, and the precision is set to $U\leq 10^{-5}$. Note that the maximum and minimum of the possible values of $U$ are completely dependent on the set of the model parameters and the configuration at that moment. In addition, each component of $\Vec{G}$ has a potential divergence due to its osmotic part: $U$ might approach positive infinity. Nonetheless, it would be helpful to show some practical value of $U$ to judge the validity of the numerical solution. In fact, when the tissue is completely packed by regular hexagons of unit area, one can estimate the maximum value of $|G_I|^2$ using the above parameters as
\bea
  |G_I|^2 \leq 3.54 , 
\eea
up to three digits. Therefore, we believe that the above precision is indeed satisfactory.

\subsubsection{Examples}

\begin{figure}[h]
  \centering
  \includegraphics[width=8.5cm]{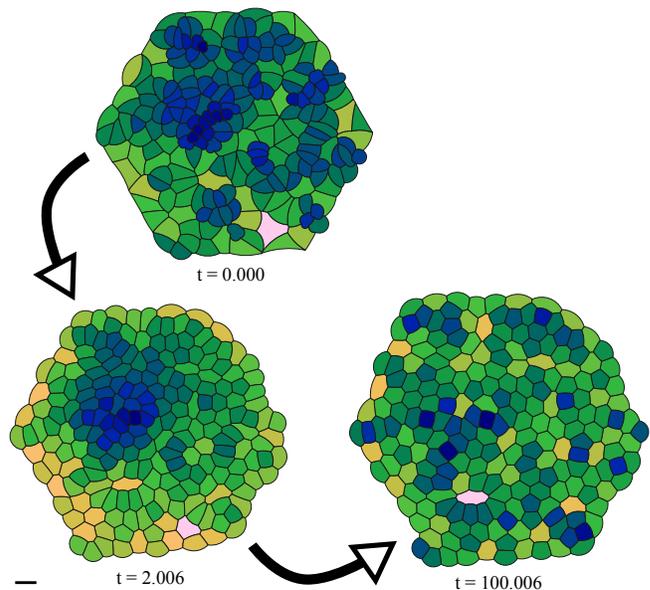}
  \caption{A series of images for a time course of the BVD generated by simulation. The images were acquired at $t=0.000$, $t=2.006$, and $t=100.006$ (from top left to bottom right). We ran the simulation up to $t \simeq 10000$, and no significant change was observed in the tissue shape after $t\simeq100$. The colors (tones) represent relative values of cell areas within an image: the darker the color is, the smaller the cell area is.}
  \label{fig:pathway}
\end{figure}

\begin{figure}[h]
  \centering
  \includegraphics[width=8.5cm]{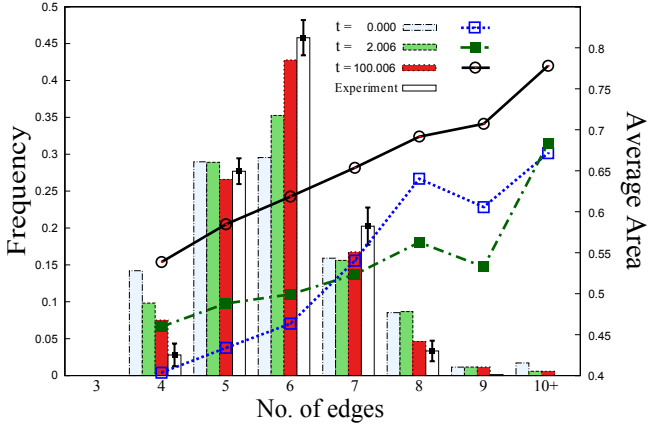}
  \caption{The clustered histograms show a time course of the frequency distribution of the cells with respect to the number of edges per cell, and the experimental observation in \cite{gibson06}. To compare with the experiment, only the non-peripheral cells are counted. From left to right in each cluster, they are of $t=0.000$, $t=2.006$, $t=100.006$, and the mean with SD of the experimental data of Drosophila imaginal discs in the late third instar stage. The line plots show a time course of the average cell areas for different numbers of edges. At $t=100.006$ (solid line with circles), the polygonal distribution approximately matches the experimental data, and the areas are roughly proportional to the number of edges per cell.}
  \label{fig:stats}
\end{figure}

The first example presents a time course of the dynamical system with the above parameters (\ref{eq:nums}, \ref{eq:params}) and the initial configuration (Fig. \ref{fig:init}). We perform the simulation from $t=0$ to $t \simeq 100$ with the above mentioned free boundary condition, and three snapshots of the time course are shown in Fig. \ref{fig:pathway}. We also show some statistics with regard to the cell sizes and the number of edges per cell in Fig. \ref{fig:stats}.

As is shown in Fig. \ref{fig:pathway}, the local inequivalences in area at the initial time are relaxed to regional inequivalences by $t=2.006$ and are sufficiently relaxed at $t=100.006$. The frequency distribution with respect to the number of edges in Fig. \ref{fig:stats} indeed indicates that the initial random pattern approaches hexagonal packing with time. Fig. \ref{fig:stats} also offers a comparison between the distributions obtained by simulation and the experimental observation by Gibson et al. \cite{gibson06}. To compare them, we have only counted the non-peripheral cells. Our distribution clearly approaches the experimental result with time. Simultaneously, the line plots of the areas in Fig. \ref{fig:stats} demonstrate that the initial random pattern in area evolves to a pattern with the approximate scaling law: $\ol{A}(n)\propto n$, where $n$ is the number of edges of a cell and $\ol{A}(n)$ is the average area of $n$-edge cells. The result implies that the state at the mechanical equilibrium obeys the so-called Lewis' law \cite{lewis28}. In our model, the pressure is a monotonically increasing function of the cell area. So, it further implies that the pressure differences do not vanish and that the nonzero curvatures remain at the equilibrium.

\begin{figure}[h]
  \centering
  \includegraphics[width=8.5cm]{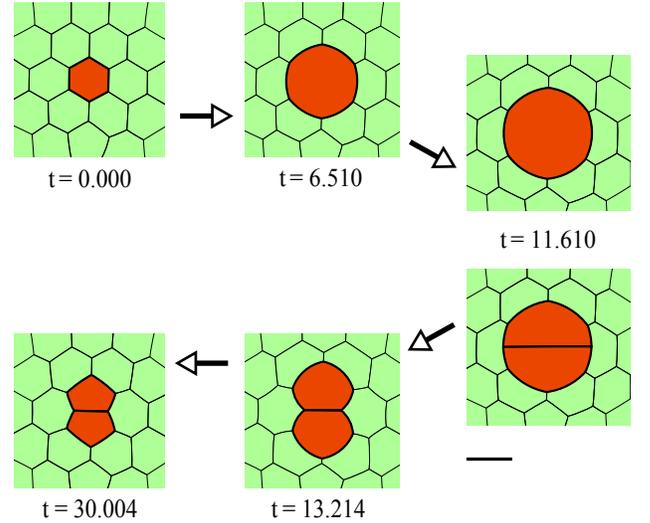}
  \caption{An example of the mitotic cell rounding during cell division obtained by simulation. The snapshots were acquired at $t=0.000,\, 6.510,\, 11.610,\, 13.214$, and $30.004$. The first three images show the extracted part of the initial configuration, the cell volume growth with the mitotic cell rounding in the $I$-phase, and the beginning of the $M$-phase. The last three images show the $M$-phase and the finalization of cytokinesis. A peanut shape appears after the cytoplasmic division at $t=11.610$.}
  \label{fig:MCR}
\end{figure}

The second example presents another time course of the system in which cell division is scheduled by the cell cycle (Fig. \ref{fig:MCR}). In eukaryotic cells, the cell cycle consists of approximately three phases: the resting phase ($G_0$-phase), interphase ($I$-phase), and mitosis ($M$-phase). The resting phase practically means the cell is not dividing, and therefore we do not need to specify any particular parameters for such cells. During interphase, the cell performs substantial work, including cell volume growth and DNA replication. During mitosis, the cell divides itself into two parts. From a mechanical point of view, the $I$- and $M$-phases could be reorganized as three stages. The first stage is the stage of cell volume growth, which is followed by the second stage of cytoplasmic division. The last stage finalizes cytokinesis. Experimentally, for example, in HeLa cells in \cite{stewart11}, cytokinesis is frequently accompanied by two mechanical phenomena around the cell boundaries: shortening of the cell boundary between a pair of daughter cells and swelling of the boundaries of the daughter cells. Thus, after inserting the edge in the second stage, during the finalizing stage, the tension of the edge shared by the daughter cells is increased, and the high internal pressure is relaxed in a smooth way to the ordinary value. 
The cell volume growth in the first stage is realized by the increase in the coefficient $R_\alpha^{(o)}$ from $1.0$ to $8.0$, which inevitably increases the target area (\ref{def:target area}) by the same magnitude. The coefficient of the cortical tension $\Gamma_\alpha^{(L)}$ is increased by a factor of $6.0$ during the growth to make the cell round. Thus, increased pressure and tension are relaxed back to the original values in the third stage. This elaborated cell cycle can in fact demonstrate the mitotic cell rounding and a peanut shape \cite{guillot13} after the division in Fig. \ref{fig:MCR}. To create the example, we first prepare a random cellular pattern in a hexagon with $N_c=61$ and relax it for $t=100$. We then switch on the cell cycle of the central cell. The central area is extracted in Fig. \ref{fig:MCR}. The model parameters are chosen to be $\Lambda_{\alpha\beta}=0.05$, $\Gamma_{\alpha}^{(L)}=0.04$, $\Gamma_{\alpha\beta}^{(l)}=0.02$, while others are the same as in the previous example, with the exception of the division-related parameter changes. The shared edge of the daughter cells has the increased parameters $\Lambda_{\alpha\beta}=0.75$, $\Gamma_{\alpha\beta}^{(l)}=0.3$ at the finalizing stage of cytokinesis.

\begin{figure}[h]
  \centering
  \includegraphics[width=6cm]{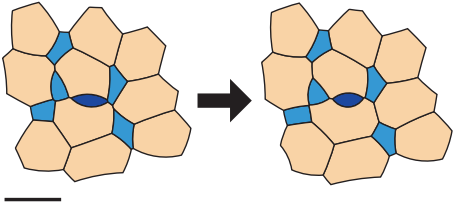}
  \caption{The dark cell at the center indicates the emerging two-vertex cell, and the other small cells share the same target area $\wt A_{minor}^{(0)} = 0.2$ and the coefficient $\Gamma_{minor}^{(L)}=0.09$. The bright large cells are of the ordinary dominant type with the target area $\wt A^{(0)} = 1.0$ and $\Gamma^{(L)}=0.03$. The two-vertex cell is surrounded by the dominant type of cells. To produce a stable two-vertex cell, we used a designed set of the parameters described in the manuscript and the initial configuration different from that in Fig. \ref{fig:init}.}
  \label{fig:tvc}
\end{figure}

\begin{figure}[h]
  \centering
  \includegraphics[width=8cm]{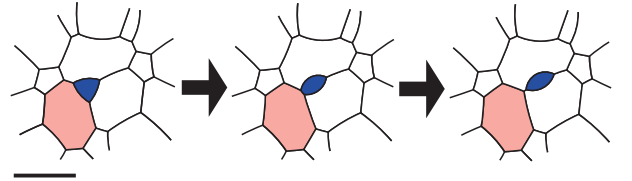}
  \caption{Successive images showing the creation of a two-vertex cell. The small, dark cell of the minor type is effectively being pushed by the light-colored, large cell of the dominant type through the repulsive interaction $\Lambda_{repulsion}=0.30$. The images were acquired at $t=0.00,\, 2.00$, and $23.60$, respectively.}
  \label{fig:createTVC}
\end{figure}

In the final examples, as a in-silico realization of the two-vertex cell which can be observed in mouse olfactory epithelium (Fig. \ref{fig:olfactory}(b)), we demonstrate the emergence of a two-vertex cell in Figs. {\ref{fig:tvc}} and \ref{fig:createTVC} through cell sorting \cite{graner92}. We emphasize here that this type of cells cannot be described by other vertex models and its demonstration proves an advantage of our model. To realize this, two types of cells are introduced, and a different initial configuration is used. First, one third of the cells in the configuration at $t\simeq 100$ in Fig. \ref{fig:pathway} are set as minor cells. Then, gradually reducing the coefficient of the osmotic pressure from $R_{minor}^{(o)}=1.0$ to $R_{minor}^{(o)}=0.2$, the cortical tension is increased by $\Gamma_{minor}^{(L)} = 3 \times \Gamma_{major}^{(L)} = 0.09$, and the adhesive interactions between major and minor cells are switched on in terms of the coefficient $\Lambda_{\alpha\beta}$ of the constant tension:
\bea
  \label{eq:adhesion}
  \Lambda_{M,M} = 0.12, \quad \Lambda_{M,m} = 0.03, \quad
  \Lambda_{m,m} = 0.21.
\eea
$M$ indicates the major cells, while $m$ indicates the minor cells. That is, the first coefficient is for the edge between major cells, the second for the relatively adhesive edge between major and minor cells, and the last for the repulsive edge between minor cells. The dominant type of cells carries the same parameters as in (\ref{eq:params}). The simplest and most naive implementation of adhesion is given by the constant shift of the tension because the adhesion is realized by the trans-bonded cadherins \cite{cadherin} and the reduction of the tension is given by its concentration in three dimensions \cite{maitre12}. Fig. \ref{fig:tvc} shows the extracted central part from the simulated tissue. The first image is a snapshot of a minor cell near the center dispatching another minor cell into the middle of the edge. The second image is after the emergence of the two-vertex cell. The difference in time between the pair of images is $\delta t=2.4$.

To confirm how easily the two-vertex cell can emerge, we have tested eight additional sets of parameters for the same initial configuration: three different sets of the interaction parameters 
\bea
  &{\rm (a)}&\Lambda_{M,M} = 0.12, \quad \Lambda_{M,m} = 0.06, \quad
  \Lambda_{m,m} = 0.18,
\nn
  &{\rm (b)}&\Lambda_{M,M} = 0.12, \quad \Lambda_{M,m} = 0.03, \quad
  \Lambda_{m,m} = 0.21,
\nn
  &{\rm (c)}&\Lambda_{M,M} = 0.12, \quad \Lambda_{M,m} = 0.00, \quad
  \Lambda_{m,m} = 0.24,
\eea
for three different $\Gamma_{minor}^{(L)}=0.06,\, 0.09,\, 0.12$. We could not have found any emergence of the two-vertex cell for the lower cortex tension $\Gamma_{minor}^{(L)}=0.06$ or for the weaker cell-cell interactions of the case ${\rm (a)}$. In other words, we have found the two-vertex cell in four cases of nine through cell sorting. This result seems to suggest that the increased cortical tension and interactions are necessary for the small cell to be a two-vertex cell. However, in a given particular configuration, not only those parameters but also the reduced target area of the cells is orchestrated for the emergence. Therefore, only some of the necessary conditions can be implied. Note that we have not found any two-vertex cell for the cases with only two types of $\Lambda$.

For the two-vertex cell to exist, the cell must have a sufficiently small target area to fit in a longer edge of larger cells. In addition, for stability, a simple physical consideration indicates that the edges connected to the two vertices of the cell are of the Hookean spring, whose tensions increase as the edges elongate. For, any small perturbation to the equilibrated vertices will be pulled back if the relevant edges are as such. We are presently unable to state a general condition for the creation of two-vertex cells. Instead, we show the situation, partially inspired by the model for metastasis \cite{brodland12}, in which the repulsive interaction effectively pushes a three-vertex cell of the minor type to the interior of an edge and creates a two-vertex cell as shown in Fig. \ref{fig:createTVC}. The repulsion is set to $\Lambda_{repulsion}=0.30$ between the minor and major cells, while the other parameters are the same as in Fig. \ref{fig:tvc}. The configuration relaxed from Fig. \ref{fig:tvc} is used as the initial configuration, and another part of the tissue is chosen for the demonstration.

\section{Discussion \& Concluding Remarks}
\label{sec:discussions}

In this paper, we have proposed and constructed the BVD, or the bubbly vertex dynamics model, as a two-dimensional geometrical and dynamical model of curved cell boundaries and (tri-) cellular junctions for monolayered epithelial tissues. As fundamental ingredients of the model, the curvatures of the cell boundaries are present and enable us to realize bubbly behaviors of the cellular dynamics, such as mitotic cell rounding and swollen cell species in epithelial tissues. 

The model of the curvatures has been formulated with a vertex model and the generalized form (\ref{def:E}) of the potential energy, where the target area arises naturally from other fundamental properties of the cells. The generalized form provides its clear physical interpretation and a variety of mechanical responses of the cells. At the characteristic time scale of epithelial tissues, the curvatures are treated in a static way, and the Young-Laplace law emerges naturally. 
To determine the curvatures, the Young-Laplace function is analyzed, which reveals that the curvatures are uniquely determined under certain simple circumstances. 
To finalize the construction, the topological changes are manifested as T1, T2, and the cell division, while T2 is redefined as T2' for the BVD. Finally, in Section \ref{sec:examples}, some examples by simulation have been presented, including a typical time course of the BVD, a realization of the mitotic cell rounding, and the emergence of the two-vertex cell, which could not be realized in the conventional vertex models. For the first example, the simple statistics of the dynamics and its brief comparison with the experimental observation in \cite{gibson06} are also given, demonstrating the approximate Lewis' law \cite{lewis28} and the enhancement of the hexagonal packing as in other geometrical models. In the second example, the cell cycle is introduced, and the peanut shape \cite{guillot13} is naturally produced as a consequence of the mitotic cell rounding and division plane shortening. Finally, a pair of particular cases are given for the emergence of the two-vertex cell whose conditions are expected to be related to the swollen cell species in epithelial tissues \cite{togashi11}.

In generalizing the potential energy, we have distinguished two typical settings for the area part of the energy, that is, the ``area constraint'' setting (AC setting: $\kappa_\alpha^{(1)}=R_\alpha^{(o)}=0$) and the osmotic pressure setting (OS setting: $\kappa_\alpha^{(2)}=0$). Although they share the same behavior near the minimum of $E_A$, they behave differently upon the application of some external force. Starting from the initial configuration in Fig. \ref{fig:init} with the parameters (\ref{eq:params}) and $P_{outer}=0.2$, we have tested the two conventional vertex models of the AC and OS settings. Our results confirm that the AC case collapses through a chain of T2 processes, while the OS case resists against the external pressure and evolves rather normally. On the other hand, the AC with $P_{outer}=0$ maintains an area similar to that of the OS. In the BVD, the AC setting gives pressure differences that are too large to yield well-defined initial curvatures for the random initial configuration, and thus the AC setting should be excluded from the BVD for general investigations.

Because the vertex models have been extensively tested and partially validated in the literature, many properties of the system realized by the BVD should not be substantially different from those of the vertex models, as long as any particular condition is imposed. To illustrate this similarity, we have tested ten random initial configurations of $217$ cells in a hexagon as in Fig. \ref{fig:init} and report that the resulting populations of $n$-sided cells and the relaxation time of the energy are similar in the three models: the AC vertex model with $P_{outer}=0$, the OS vertex model, and the BVD of the OS setting, while other parameters are homogeneously given in Eq. (\ref{eq:params}). All of the models exhibit a tendency toward hexagonal packing and approximately obey the Lewis' law.

As mentioned above, a clear difference between the vertex models and the BVD may arise in the forces acting on the cellular junctions because the pressures directly move the vertices in the former while only the tensions move them in the BVD. Therefore, if the curvatures and the pressure differences are not negligible, the resulting dynamics must be different at least locally. However, due to a lack of a clear manifestation of such a difference other than cell shapes, we have only demonstrated a local topological difference in the mitotic cell rounding and the emergence of the two-vertex cell in the examples. There are a few methods which quantitatively characterize patterns, for example, by particle image velocimetry (PIV) \cite{raffel07} or by the texture matrix \cite{graner08}. These methods might describe the difference between the BVD and other geometrical models. However, PIV and the matrices neglect the details of the cell boundaries such as the curvatures and thus would be unable to fully characterize the BVD.

Before discussing the applicability of the model, we will discuss the assumptions and the limitations of our construction. Their removals will lead to some possible extensions of the model.

On the degree of the vertices, we can relax the assumption to include vertices with degree two, which are the virtual vertices in the interior of the edges. These vertices at the boundary of the tissue can characterize different boundaries with different boundary conditions, enabling the attachment of the tissue to a firm material or the application of external forces through the boundary. This is a very useful tool for mechanical in silico experiments. The vertices with degree greater than three can achieve more complex structures, including the rosette structure \cite{blankenship06}. Practically speaking, future studies should examine those vertices with degree four for their applicability to achieve the desired phenomena. We have assumed non-negative tension primarily to prevent the buckling of the cell boundaries. One can include the buckling with the vertices with degree two, but this may yield multiple solutions of the law (\ref{def:YLlaw}) for numerical calculations. 

We have excluded the bending energy, the angular energy, and the chemical potential. The bending rigidity of the cell boundary is considered negligible relative to the size of the cell boundaries. However, the bending rigidity may acquire non-negligible magnitude if the actomyosin complex accumulates to extraordinary levels at some edge or some other stiffer cytoskeleton is attached to the cell boundary. To generalize our model for such situations, the curvatures should be replaced by some elliptic integrals to express the edges \cite{landau}. This generalization is also capable of dealing with both the buckling effect and the collisions of the cell boundaries, and thus represents an interesting direction of study. While the implementation of the angular energy is relatively easy, we leave this possibility for those who consider it necessary around the cellular junctions. We have not included the chemical potential to focus on the pure mechanical properties of the tissues. It would be intriguing to consider chemical reactions and environment of several types of adhesive proteins, such as cadherins and nectins, or edge-related proteins such as actin fibers and myosins. Such studies would represent the next step for the application of the model to the real phenomena in tissues.

The dynamical degrees of freedom of the curvatures can be recovered if we replace the evaluation function $U$ with the energy $E$. The method of steepest descent for a step at a time is equivalent to Euler's method for the dynamics of curvatures. Unfortunately, the information needed to determine the friction term of the curvature is not available. If one knows or is able to infer this information, the extension of the current construction is straightforward. Alternatively, the Surface Evolver \cite{brakke92} or Kawasaki's model \cite{kawasaki89} with elaborate friction coefficients and energy might be more appropriate for the simulation of the dynamical curvatures. Any of the above three methods would serve as novel tools to investigate the dynamics of epithelial tissues with frequently changing cell boundaries or rapidly changing tensions.

The saturation bounds of the edges have been introduced so that the overlapping or colliding edges are mostly avoided. However, even with the bounds, a random initial configuration with overlapping edges occasionally occurs. Therefore, it would be ideal to remove the bounds and introduce the cell boundary collision instead. The removal of the bounds also requires another parameterization of the curved edges, because the shape of the edge becomes a double-valued function of the curvature radius in the case without the bounds. This generalization may be related to that for the bending energy if we treat the cell area as an elastic body, and would be an ambitious future work.

In addition to the possible extensions related to the assumptions and limitations, there are a variety of other possibilities for the model. For example, one may ask how the model can be generalized to three-dimensional cases. If we assume the Young-Laplace law and a uniform tension for a cell boundary, then we can assign a single mean curvature to the cell boundary. Therefore, the simplest and most naive extension would result in solving the boundary problems of the cell boundaries for given mean curvatures, uniform tensions, and vertex positions. Achieving this solution is not at all trivial and would be a meaningful direction for the generalization of this model. Another interesting direction is the simulation methods. The model can be applied to a Monte-Carlo simulation by replacing the dynamical portion with the Metropolis-Hastings algorithm or to an event-driven simulation with energy minimization \cite{farhadifar07,hufnagel07}.

We have discussed the construction and possible extensions of our model rather than restricting our discussion to a complete characterization of BVD. For this characterization, it would be intriguing to investigate the effect of the nonzero value of the buffering area $V_\alpha$. Its presence would be critical when the external force or the pressure differences are large enough to push neighboring cells out of the plane through T2'. The latter condition may arise when the cell divisions or local tissue growth dominantly determine the tissue dynamics, as suggested in \cite{shraiman05} as the ``cell competition''. Second, the ground state of the system has not been fully investigated in the BVD. In \cite{farhadifar10}, the ground state of the vertex model with the energy (\ref{def:Evdm}) is analyzed based on the stability analysis of a single polygonal cell. Unfortunately, in our model, the cells do not take simple polygonal shapes in general, and we cannot expect the validity of the single-cell analysis. Therefore, the ground-state analysis should be generalized first from that in \cite{farhadifar10} for the BVD and its ``phase'' diagram. Third, stochastic noise and the referential position may be taken into account in the EOMs (\ref{def:EOMs}). The vertices are the ``fictitious'' physical objects, and we are therefore hesitant to implement the noise in the naive way. We infer that the noise should be attributed to the mechanical properties of the tissue rather than directly to the EOMs, as the annealing procedure in \cite{farhadifar07}. In addition, it would be interesting to investigate the effect of the nonzero referential position $\Vec{X}_i$. Because the position as an averaged quantity of the neighborhoods can be interpreted as a finite expression of the viscous property of the tissue, it will lead to nontrivial differences in tissue dynamics, as suggested for wound healing in \cite{cochet14} and three-dimensional tubular formation in \cite{okuda14}.

In principle, our model can be extended in the same manner that the vertex models have been extended, and is applicable to all of their applications, ranging from fundamental studies on packing and patterns to in silico experiments such as laser ablations and growing tissues. In particular, in silico experiments with external forces would be interesting because the forces acting on the vertices differ between the BVD and the vertex models. We anticipate that an extensive simulation study of BVD would reveal some critical differences with other geometrical models. As emphasized in the introduction, the BVD can be applied to tissues with highly curved cells, such as mouse auditory \cite{togashi11} or olfactory epithelia, or Drosophila retinal epithelia, as long as the two-dimensional description of these tissues is acceptable. The arrangement of the different cell species or, for short, the cell sorting in these epithelia is likely to be deeply related to the relevant gene expressions and mechanically associated proteins, such as adhesive proteins, elastic proteins of actomyosin complexes, and their regulators. Therefore, for such applications, the model must be equipped with relevant biochemical reactions and interactions. Another type of application is the cell competition; Shraiman proposed in \cite{shraiman05} that cell death in the competition may be controlled by local differences in growth rates. The difference is equivalent to the pressure difference in his description, and our model is one of the best discrete models to describe this difference. These are all exciting phenomena for application of the BVD.

\vspace*{10pt}
\noindent 
{\bf Acknowledgments}

The authors are grateful to Prof. H. Honda, Prof. F. Graner, Dr. P. Marcq, Dr. K. Sato, Prof. S. Ishihara, Dr. K. Sugimura, and Dr. H. Togashi for their stimulating discussions and suggestions. The authors are grateful to Prof. Y. Sasai for his continuous encouragement. The authors would also like to acknowledge all of the members of our laboratory for developmental morphogeometry for their continuous support and stimulating discussions. One of the authors (YI) would like to thank people at Physical Chemistry Curie and Genetics and Developmental Biology, Institut Curie, for their hospitality and stimulating discussions. YI would also like to thank N. Ishimoto for his continuous encouragement. This work is supported by JSPS KAKENHI (Grants-in-Aid for Scientific Research) Grant Number 26540158.

\appendix
\vspace{20pt} \noindent {\Large\bf Appendix}
\section{The non-dimensionalization}
\label{app:nondimensionalization}

The energy (\ref{def:E}) can be non-dimensionalized with an arbitrary constant area $A_0$, and it is convenient to use the average area of the cells. First, the areas $A_\alpha$ and $V_\alpha$ are normalized by $A_0$. Then, the coefficient of the surface tension is factored out, and the coefficients are renormalized in the expression (\ref{def:E}). Namely,
\bea
\label{eq:nondimensionalisation-app}
  E
   &=& \sum_{\braket{i,j}} \Lambda_{\alpha \beta}\, l_{ij}
    + \sum_{\braket{i,j}} \frac{\Gamma_{\alpha\beta}^{(l)} }{2} (l_{ij})^2
\nn&&
    + \sum_\alpha \Biggl[ 
    \kappa_\alpha^{(1)} A_0 \frac{A_\alpha}{A_0} 
    -  R_\alpha^{(o)} \log \left( \frac{A_\alpha}{A_0} + \frac{V_\alpha}{A_0} \right) 
\nn&&\qquad
    + \frac{\Gamma_{\alpha}^{(L)} A_0}{2} \frac{L_{\alpha}^2}{A_0}
    \Biggr]
\nn
  &=& \sum_\alpha \wh \kappa_\alpha \Biggl[
     \wh A_\alpha
    -  \wh R_\alpha^{(o)} \log \left( \wh A_\alpha + \wh V_\alpha \right) 
    + \frac{\wh\Gamma_{\alpha}^{(L)} }{2} \wh L_{\alpha}^2
\nn&&\qquad
    + \sum_{\braket{i,j}\in\alpha} \left\{ \frac{\wh \Lambda_{\alpha \beta}}{2}\, \wh l_{ij}
    + \frac{\wh \Gamma_{\alpha\beta}^{(l)} }{4} (\wh l_{ij})^2 \right\}
   \Biggr] ,
\nn
\eea
up to a constant, where $\wh \kappa_\alpha \equiv \kappa_\alpha^{(1)}{A_0}$. The relationship between the renormalized quantities and the original quantities is
\bea
\label{def:renormalisation}
  \wh A_\alpha &=& \frac{A_\alpha}{A_0}, \quad
  \wh V_\alpha \,=\, \frac{V_\alpha}{A_0},\quad
\nn
  \wh L_\alpha &=& \frac{L_\alpha}{\sqrt{A_0}}, \quad
  \wh l_{ij \in \alpha} \,=\, \frac{l_{ij}}{\sqrt{A_0}}, 
\nn
  \wh R_\alpha^{(o)} &=& \frac{R_\alpha^{(o)}}{\kappa_\alpha^{(1)} A_0},\quad
  \wh \Gamma_\alpha^{(L)} \,=\, \frac{\Gamma_{\alpha}^{(L)}}{\kappa_\alpha^{(1)}}, 
\nn 
  \wh \Lambda_{\alpha\beta} &=& \frac{\Lambda_{\alpha\beta}}{\kappa_\alpha^{(1)} \sqrt{A_0}},\quad
  \wh \Gamma_\alpha^{(l)} \,=\, \frac{\Gamma_{\alpha}^{(l)}}{\kappa_\alpha^{(1)}}.
\eea
From the first line of Eq. (\ref{eq:nondimensionalisation-app}) to the second, we distributed the contribution of the edge lengths, $\{l_{ij}\}$, to the neighboring cells that share the edges. The rescaled target area $\wh A_\alpha^{(0)}\equiv \wt A_\alpha^{(0)}/ A_0$ is given by substituting Eq. (\ref{def:renormalisation}) into Eq. (\ref{def:target area}):
\bea
\label{rel:VR}
  \wh A_\alpha^{(0)} = - \wh V_\alpha + \wh R_\alpha^{(o)}.
\eea
In particular, when $V_\alpha=0$, the target area is equal to the osmotic constant $\wh R_\alpha^{(o)}$. If there is a dominant cell type sharing the same target area $a$, then we suggest that the dimensionful variables be normalized by $a$ of the dominant type. This normalization enables some target areas to remain different from unity for minor cell types. The relation (\ref{rel:VR}) takes the same form for a minor cell type, while that of the dominant cell type is normalized to unity.

Finally, we comment on the $\kappa_\alpha^{(1)}=0$ case without boundary. In the absence of osmotic pressure, $A_\alpha^{(0)}$ is the target area, as shown. In the presence of osmotic pressure, the minimum of the area energy of a single cell also depends on the values of $R_\alpha^{(o)}$ and $V_\alpha$. In fact, the target area is given by 
\bea
\wt A_\alpha^{(0)} &=& \frac12 \left( A_\alpha^{(0)}-V_\alpha 
 + \sqrt{(A_\alpha^{(0)}+V_\alpha)^2 + 4 R_\alpha^{(o)}/\kappa_\alpha^{(2)}} \right)
\nn 
  &\simeq& A_\alpha^{(0)} + \frac{R_\alpha^{(o)}}{\kappa_\alpha^{(2)} (A_\alpha^{(0)}+V_\alpha)}
 > A_\alpha^{(0)}, 
\eea
where $R_\alpha^{(0)}/\kappa_\alpha^{(2)} \ll (A_\alpha^{(0)}+V_\alpha)$ is assumed for the last approximation. This is still relatively complicated, and thus we suggest performing the non-dimensionalization by the original ``target area'' rather than the precise target area given above.

\section{The derivation of the resummation formula}
\label{app:pdA}

In this section, we derive the resummation formula:
\bea
\label{formula:resummation}
  \sum_\alpha p_\alpha \delta A_\alpha = \sum_I (p_{\alpha_I} - p_{\beta_I}) \delta A_{\alpha_I, I}, 
\eea
where we use the same conventions as in the section \ref{sec:minimisation}. We also intend to show how to decompose the $I$-th edge fraction of the variation of the area, $\delta A_{\alpha_I,I}$, into three parts:
\bea
  \delta A_{\alpha,I} 
  &=& R_I \delta l_I
    + R_I \sin\theta_I \delta l_I^{\perp} 
    - R_I \cos\theta_I \delta l_I^{(0)} .
\eea

The area $A_\alpha$ consists of the polygonal area $A_\alpha^{(poly)}$ and the sum of the comb shapes associated with the edges as in Eq. (\ref{def:A}).
\bea
  A_\alpha &=& A_\alpha^{(poly)} + \sum_{I \in \alpha} (\sgn\!{}_I \alpha) A_I^{(comb)} ,
\eea
where $A_I^{(comb)} \equiv R_I^2 \left\{ \theta_I - \frac12 \sin (2\theta_I) \right\}$. The symbol $(\sgn\!{}_I \alpha)$ is introduced to reflect the direction of the $I$-th edge against the cell $\alpha$: $\sgn\!{}_I \alpha = 1$ if $\alpha$ is the left cell to the $I$-th edge, and $\sgn\!{}_I \alpha = -1$ if $\alpha$ is the right. The variation of the area is simply given by $\delta A_\alpha = \delta A_\alpha^{(poly)} + \sum_{I\in \alpha} (\sgn\!{}_I \alpha) \delta A_I^{(comb)}$ because the symbol remains the same with the changes in $\{\x_i\}$ and $\{R_I\}$. The variation of the comb shape $\delta A_I^{(comb)}$ is naturally attributed to the $I$-th edge, and thus the problem is reduced to the decomposition of $\{\delta A_\alpha^{(poly)}\}$.

\begin{figure}[h]
  \centering
  \includegraphics[width=5cm]{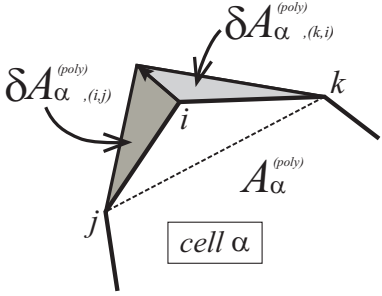}
  \caption{The $\delta \x_i$ contribution in $\delta A_\alpha^{(poly)}$.}
  \label{fig:triangle}
\end{figure}

Recall that the variations are primarily given by those of the vertex positions and the curvatures. Since the polygonal areas do not depend on the curvatures, the variations $\{\delta A_\alpha^{(poly)}\}$ can be rewritten solely by $\{\delta \x_i\}$. Let us select a vertex $\x_i$ of the cell $\alpha$ and consider $\delta A_\alpha^{(poly)}$ under the variation $\delta \x_i$ only. The relevant change of $\delta A_\alpha^{(poly)}$ under the change $\delta \x_i$ is that of the triangular area of the vertices $i, j$, and $k$, for $j$ and $k$ adjacent to $i$ (Fig. \ref{fig:triangle}). Then, $\delta A_\alpha^{(poly)}$ can be expressed by
\bea
  \delta A_\alpha^{(poly)} &=& \frac12 \delta \left\{ (\x_j - \x_i) \times (\x_k - \x_i) \right\}_z
\nn
  &=& \frac12 \left\{ \delta \x_i \times (\x_i - \x_k) \right\}_z
     + \frac12 \left\{ \delta \x_i \times (\x_j - \x_i) \right\}_z
\nn
  &=& \frac12 l_{ki}^{(0)} \delta l_{ki}^{\perp}
     + \frac12 l_{ij}^{(0)} \delta l_{ij}^{\perp} , 
\eea
where $\times$ denotes the cross product of the vectors. We temporarily extended the two-vectors to three-vectors and extracted the $z$-components of them. $\delta l_{ki}^{\perp}$ denotes the component of $\delta \x_i$ perpendicular to the edge $(k,i)$, pointing right to the edge $(k,i)$. 
Therefore, it is evident that $\delta A_\alpha$ can be decomposed into its edge fractions:
\bea
\label{eq:edgefraction}
  \delta A_\alpha = \sum_{I \in \alpha} \delta A_{\alpha,I}.
\eea
Since the cells are densely packed in the tissue, the $I$-th edge fraction of $\delta A_\alpha$ is the same as the $I$-th edge fraction of $\delta A_\beta$ when they share the $I$-th edge, except for its signature. Namely,
\bea
\label{eq:edgefraction-identity}
  \delta A_{\alpha, I } = - \delta A_{\beta,I}.
\eea
The formula (\ref{formula:resummation}) can be confirmed by substituting Eqs. (\ref{eq:edgefraction}, \ref{eq:edgefraction-identity}) in the left hand side.

The derivation of the formula demonstrates that the $I$-th edge fraction of the variation of the area can be written explicitly under the change $\delta R_I$ and the change $\delta \x_i$ of the $i$-th vertex belonging to the $I$-th edge:
\bea
&&  \delta A_{\alpha,I} 
\nn&&
  = \frac12 l_I^{(0)} \delta l_I^{\perp} 
    + \delta \left\{ R_I^2 \theta_I - \frac12 R_I^2 \sin(2\theta_I) \right\}
\nn&&
  = \frac12 l_I^{(0)} \delta l_I^{\perp} 
    + \left\{ 2 R_I \theta_I \delta R_I + R_I^2 \delta \theta_I
    - \frac12 \delta\left( R_I^2 \sin(2\theta_I) \right) \right\}
\nn&&
  = R_I \delta l_I
    + \frac12 l_I^{(0)} \delta l_I^{\perp}
    - \cos\theta_I \left\{ l_I^{(0)} \delta R_I
       + 2 R_I^2 \cos\theta_I \delta\theta_I \right\}
\nn&&
  = R_I \delta l_I
    + R_I \sin\theta_I \delta l_I^{\perp} 
    - R_I \cos\theta_I \delta l_I^{(0)} .
\eea
From the second line to the third, the identity $\delta l_I = 2\left( \theta_I \delta R_I + R_I \delta \theta_I \right)$ is used. From the third to the last, another identity, $\delta(\sin\theta_I) = \delta(l_I^{(0)}/(2R_I))$, is used.

\section{Proofs and limitations on the solution of the Young-Laplace equations}
\label{app:proofs}

In this section, we prove the uniqueness of the solutions of the Young-Laplace equations in its analytic domain. To be precise, we state that, for the Young-Laplace function $\Vec{G}(\Vec{\rho})$ defined by Eqs. (\ref{eq:Ggeneral}, \ref{def:Gmodification}), 
\begin{proposition}
\label{proposition}
if there is a solution of Eq. (\ref{def:vYL}) in the analytic domain of $\Vec{G}$ and if the parameters satisfy the following condition, the solution is isolated and unique in the analytic domain.
\end{proposition}
The condition and, consequently, the limitation of the above are given by
\bea
\label{cond:proofs}
  B_\alpha(A_\alpha) l_I^{(0)} l_J^{(0)} \geq 4 \Gamma_\alpha^{(L)} 
  \quad {\rm for~any~}\alpha {\rm ~and~} I,J \in \alpha ,
\eea
where $B_\alpha(A_\alpha) \equiv \kappa_\alpha^{(2)} + R_\alpha^{(o)}/(A_\alpha+V_\alpha)^2$ with $\kappa_\alpha^{(2)}>0$ or $R_\alpha^{(o)}>0$. The definition of the analytic domain is given in Eq. (\ref{def:analytic domain}) in the following section. Notably, when $\Gamma_\alpha^{(L)}=0$ for any $\alpha$, the condition is trivially met because all coefficients in the condition are positive semidefinite. Or, for ordinary, almost hexagonally packed cells, $A_\alpha$ and, consequently, $B_\alpha$ and $l_{I,J}^{(0)}$ are of order $O(1)$, so that a small value of $\Gamma_\alpha^{(L)}$ is sufficient. Note that, by the definition (\ref{eq:Ggeneral}), non-negative tension is supposed, {\it i.e.}, $\ol \Gamma_{\alpha\beta} l_I^{(0)} \geq - \ol \Lambda_{\alpha\beta}$.

In the following, as is stated above, we assume the existence of at least one root in the analytic domain of the function $\Vec{G}$. The function $\Vec{G}$ is not polynomial but analytic in its domain except for the discontinuously saturated points. Therefore, the space of the solutions of Eq. (\ref{def:vYL}) would not be an algebraic but an analytic variety if it exists. In such a case, it is highly nontrivial to show the existence of the solutions. Therefore, we simply assume the existence of the solutions, upon which we reveal their nature. By clarifying some features of the domain and the analyticity of the function, we prove Proposition \ref{proposition} step by step. 

\subsection{On the domain of the Young-Laplace functions}
\label{app:domain}

Let us focus on the domain of the function $\Vec{G}$ defined by Eqs. (\ref{eq:Ggeneral}, \ref{def:Gmodification}). We use the normalized variables $\{\rho_I\}$ in Eq. (\ref{def:normalised-variable}) to describe the domain. Due to the saturation bounds of the curvatures, $\rho_I$ is in the closed interval $D\equiv [-1,1]$ by definition. Therefore, the domain of the function is, at most, the $N_e$-dimensional hypercube $D^{N_e}$ for $N_e$ edges. In other words, the function $G_I$ is expected to map $D^{N_e}$ to $\R$, and the implicit equation $G_I = 0$ prescribes a $(N_e-1)$-dimensional hypersurface in $D^{N_e}$. However, the domain is actually a subspace of $D^{N_e}$ and must be modified for our root finding problem, $\Vec{G}=0$, according to the analytic behavior of the function $\Vec{G}$.

A decrease of a subdomain of $D^{N_e}$ may arise either from a requirement of non-negative cell areas or from an analytic treatment of the singular terms in the expression (\ref{eq:Ggeneral}). For clarity, we first describe the function below and then discuss the decreases. 

The function is defined for the given $\{\x_i\}$ and the parameters, which appear in the expression as
\bea
\{l_I^{(0)}\}, \{A_\alpha^{(poly)}\}, 
\{ \kappa_\alpha^{(1,2)}\}, \{\Lambda_{\alpha\beta}\}, \{\Gamma_{\alpha\beta}^{(l)}\}, 
\nn
\{\Gamma_\alpha^{(L)}\}, \{R_\alpha^{(o)}\}, \{A_\alpha^{(0)}\}, \{V_\alpha\}, P_{outer}.
\eea
The comb areas $\{A_I^{(comb)}\}$ are the functions of $\{\rho_I\}$ and should be treated separately when analyzing the function. Namely, the $I$-th component of the function, $G_I$, can be viewed as a function of $\rho_I$ for fixed $\{\rho_{J\neq I}\}$. Decomposing the areas according to their $\rho_I$-dependencies, one finds: 
\bea
\label{eq:Gnormalised2}
&&
  G_I(\rho_I;{\rho_{J\neq I}})
\nn&&
  = 2 
  \left( \ol \Gamma_{\alpha\beta} \arcsin\rho_I 
    + \rho_I \ol \Lambda_{\alpha\beta}/l_I^{(0)} \right)  
\nn&&\quad
  + \left( \kappa_\alpha^{(2)} + \kappa_\beta^{(2)} \right)  A_I^{(comb)}
\nn&&\quad
  {}- \frac{ R_\alpha^{(o)} }{ A_I^{(comb)} + V_\alpha +A_\alpha|_{\rho_I=0} }
  + \frac{ R_\beta^{(o)} }{ - A_I^{(comb)} + V_{\beta} +A_\beta|_{\rho_I=0}} 
\nn&&\quad
  {}- P_{const}
  + \left.\left( \kappa_\alpha^{(2)} A_\alpha - \kappa_\beta^{(2)} A_\beta \right)\right|_{\rho_I=0} ,
\eea
where $\alpha$ and $\beta$ denote the cells to the left and right of the $I$-th edge. The comb area $A_I^{(comb)}$ is
\bea
  A_I^{(comb)} (\rho_I)
  &=& \frac{(l_I^{(0)})^2}{4 \rho_I^2} \left\{
   \arcsin \rho_I
   - \rho_I \sqrt{1- \rho_I^2 }
  \right\} .
\eea
From above, it is trivial that $A_I^{\!(\!comb)}\!(1)=-A_I^{\!(\!comb)}\!(-1)=\pi (l_I^{(0)})^2 / 8$. Because it can be shown that the comb area is a monotonic function of $\rho_I$, {\it i.e.}, $\frac{\pa}{\pa \rho_I} A_I^{(comb)} > 0$ for $\rho_I \in D$, the following inequalities hold for $\rho_I \in D$:
\bea
\label{eq:comb-bounds}
-\frac{\pi (l_I^{(0)})^2}{8} \leq A_I^{(comb)}(\rho_I) \leq \frac{\pi (l_I^{(0)})^2}{8}.
\eea

If one requires positive semidefiniteness of the cell areas, then the comb area and its associated cell areas must satisfy the following inequalities:
\bea
\label{eq:area-ineq}
 -A_\alpha |_{\rho_I=0} \leq  A_I^{(comb)}(\rho_I) \leq  A_\beta |_{\rho_I=0} .
\eea
Furthermore, because of the bounds (\ref{eq:comb-bounds}), the following two statements hold:
\bea
&{\rm (i)}&  {\rm if~} \left\Vert A_\alpha |_{\rho_I=0} \right\Vert \leq \frac{\pi (l_I^{(0)})^2}{8}, 
\nn&&
  {}^\exists \rho_I^- {\rm ~s.t.~} -1\leq \rho_I^- \leq 1 {\rm ~and~}
  A_I^{(comb)}(\rho_I^-) = - A_\alpha |_{\rho_I=0}.
\nn
&{\rm (ii)}&  {\rm if~} \left\Vert A_\beta |_{\rho_I=0} \right\Vert \leq \frac{\pi (l_I^{(0)})^2}{8}, 
\nn&&
  {}^\exists \rho_I^+ {\rm ~s.t.~} -1\leq \rho_I^+ \leq 1 {\rm ~and~}
  A_I^{(comb)}(\rho_I^+) = A_\beta |_{\rho_I=0}.
\nn
\eea
Here, $\|*\|$ denotes the absolute value for convenience. Therefore, if the conditions of (i) or (ii) are satisfied, the interval $D$ for $\rho_I$ should be modified to $\wt D$, such that
\bea
\label{def:domain1}
  \wt D &=& [\rho_I^-,1] \quad {\rm if~ the~ condition~ (i) ~holds},
\nn
  {\rm or~} \wt D &=& [-1,\rho_I^+] \quad {\rm if~ the~condition~ (ii) ~holds},
\nn
  {\rm or~} \wt D &=& [\rho_I^-,\rho_I^+] \quad {\rm if~ the~conditions~ (i,ii) ~hold}.
\eea
From the inequality (\ref{eq:area-ineq}), $\rho_I^- \leq \rho_I^+$ and $\wt D \neq \emptyset$. Note that the areas, $A_\alpha|_{\rho_I=0}$ and $A_\beta|_{\rho_I=0}$, contain not only the polygonal areas independent of $\Vec{\rho}$ but also the comb areas of other edges, and thus $\rho_I^\pm$ depends on the values of $\{\rho_{J \neq I}\}$ for $J \in \alpha,\beta$ under the conditions of (i, ii). This dependence defines nontrivial $(N_e-1)$-dimensional hypersurfaces as the boundaries along the transverse direction to the $\rho_I$ axis.

Formally, there is another possibility of decreasing a subdomain of $D^{N_e}$. It is obvious in the expression (\ref{eq:Gnormalised2}) that there are two potential singularities originating from the osmotic pressure. In fact, because of the inequalities (\ref{eq:comb-bounds}), $G_I$ becomes divergent at $\rho_I = \varrho_I^-$ or $\varrho_I^+$, if the denominators of the pressures satisfy the following inequalities:
\bea
\label{def:singularities}
&{\rm (iii)}&  {\rm if~} \left\Vert V_\alpha+A_\alpha |_{\rho_I=0} \right\Vert
                   \leq \frac{\pi (l_I^{(0)})^2}{8}, 
\nn&&
  {}^\exists \varrho_I^- {\rm ~s.t.~} -1\leq \varrho_I^- \leq 1 
\nn&&\qquad
{\rm ~and~}
  A_I^{(comb)}(\varrho_I^-) = - V_\alpha - A_\alpha |_{\rho_I=0}.
\nn
&{\rm (iv)}&  {\rm if~} \left\Vert V_{\beta} + A_\beta |_{\rho_I=0} \right\Vert \leq \frac{\pi (l_I^{(0)})^2}{8},
\nn&& 
  {}^\exists \varrho_I^+ {\rm ~s.t.~} -1\leq \varrho_I^+ \leq 1 
\nn&&\qquad
{\rm ~and~}
  A_I^{(comb)}(\varrho_I^+) = V_{\beta} + A_\beta |_{\rho_I=0}.
\nn
\eea
Correspondingly, the interval is subject to the modification
\bea
\label{def:domain2}
  \wt D &=& (\varrho_I^-,1] \quad {\rm if~ the~ condition~ (iii) ~holds},
\nn
  {\rm or~} \wt D &=& [-1,\varrho_I^+) \quad {\rm if~ the~condition~ (iv) ~holds},
\nn
  {\rm or~} \wt D &=& (\varrho_I^-,\varrho_I^+) \quad {\rm if~ the~conditions~ (iii,iv) ~hold}.
\nn
\eea
However, it is not difficult to find that $\varrho_I^\pm$ and $\rho_I^\pm$ obey the following inequality due to the non-negative constant $V_\alpha\geq 0$:
\bea
  \varrho_I^- \leq \rho_I^- \quad {\rm and} \quad 
  \rho_I^+ \leq \varrho_I^+.
\eea
The equalities hold when $V_\alpha=0$. In such a situation, the domain (\ref{def:domain2}) is given a priority over the domain (\ref{def:domain1}) and defines the domain of the function $\Vec{G}$, excluding the boundary points as above. By contrast, when $V_\alpha>0$, the domain (\ref{def:domain1}) defines that of the function. Note $\wt D \subseteq D$.

Hence, setting $\wt D = D$ for trivial edges, the domain of the function $\Vec{G}$ can be expressed as 
\bea
\label{def:domain}
\wt D^{N_e} \subseteq D^{N_e},
\eea
defined as above. 

The function $\Vec{G}$ is defined on the entire domain $\wt D^{N_e}$, but it is not necessarily smooth or continuous on the entire domain due to the modification (\ref{def:Gmodification}). That is to say, we define the determination of the curvatures as the root finding problem of $G_I$s. Accordingly, $G_I$ was forced to be zero at the end of $\wt D$ for $\rho_I$, when $G_I$ takes only negative, or positive, values in $\wt D$. The set of such discontinuously saturated points on $\wt D^{N_e}$ forms a subdomain of $\wt D^{N_e}$ and is located at the boundary of the domain. We call it the special boundary defined by the modification and label it by $\pa_{dis} \wt D^{N_e}$. Taking the set difference, we find the analytic domain of the function $\Vec{G}$ as
\bea
\label{def:analytic domain}
  \wt D^{N_e} \backslash \pa_{dis} \wt D^{N_e}.
\eea
In other words, the function is analytic on $\wt D^{N_e}$ except for the special boundary $\pa_{dis} \wt D^{N_e}$. In an approximate but technical sense, the special boundary constitutes, at most, $2N_e$ disconnected $(N_e-1)$-dimensional subdomains of $\wt D^{N_e}$, each of which is an analytic subdomain at the boundary of the interval for an edge. Therefore, by fusing them with the analytic domain, one can consider the analytically continued domain or the extended analytic domain of the function, on which the function is analytic. This space is analogous to the complex plane with a branch cut. Therefore, the function may behave well on this extended domain.

\subsection{On the monotonicity of the Young-Laplace functions and the Jacobian matrix}
\label{app:monotonicity}

On the domain $\wt D^{N_e}$, the unmodified form of the function $G_I$ turns out to be a monotonically increasing function of $\rho_I$. In addition, each non-vanishing component of the Jacobian matrix is found to be definite if the condition (\ref{cond:proofs}) holds. These two statements play key roles in proving Proposition \ref{proposition}. 

On $\wt D^{N_e}$, it is obvious that the unmodified form of the function is analytic, {\it i.e.}, continuous and smooth. Therefore, to show that it is a monotonically increasing function of $\rho_I$, it is sufficient to demonstrate that its first derivative in $\rho_I$ is positive semidefinite in the interval $(\min(\rho_I),\max(\rho_I))$. The first derivative can be explicitly written as
\bea
\label{eq:G'}
&&
  \frac{\pa G_I}{\pa \rho_I }
\nn&&
  = \frac{2 \ol \Gamma_{\alpha\beta}}{\sqrt{1-\rho_I^2}} 
  + \frac{2 \ol \Lambda_{\alpha\beta}}{l_I^{(0)} }
\nn&&\quad
  + \left(
    \kappa_\alpha^{(2)} + \kappa_\beta^{(2)} 
    + \frac{R_\alpha^{(o)}}{ \left( A_{\alpha}+V_\alpha \right)^2 }
    + \frac{R_\beta^{(o)}}{ \left( A_{\beta }+V_{\beta} \right)^2 }
    \right)
  \frac{\pa A_I^{(comb)} }{\pa \rho_I} 
\nn&&
  = \frac{2 \ol \Gamma_{\alpha\beta}}{\sqrt{1-\rho_I^2}}
  + \frac{2 \ol \Lambda_{\alpha\beta}}{l_I^{(0)}} 
\nn&&\quad
  + \frac{(l_I^{(0)})^2}{4} \left( B_\alpha + B_\beta \right) 
    \left( \frac{1}{ \sqrt{1-\rho_I^2}} + 2 F(\rho_I)\right) ,
\eea
where $B_\alpha$ is the same as in the condition (\ref{cond:proofs}), and $F(\rho_I)$ denotes 
\bea
 F(\rho_I ) &\equiv& 
  \frac{1}{\rho_I^2}
  \left( \frac{1}{\sqrt{1-\rho_I^2}} - \frac{\arcsin \rho_I}{\rho_I} \right), 
\eea
and we have used the relation
\bea
  \frac{\pa A_I^{(comb)}}{\pa x_I}
  &=& \frac{(l_I^{(0)})^2}{4} \left( 
       \frac{1}{\sqrt{1-\rho_I^2}} + 2 F(\rho_I) \right).
\eea
Because of the condition of the non-negative tension $T_I\geq 0$, {\it i.e.}, $\ol \Gamma_{\alpha \beta} l_I^{(0)} + \ol \Lambda_{\alpha \beta} \geq 0$, the first two terms in (\ref{eq:G'}) are non-negative. The first two factors of the last term are positive. The positivity of $F(x)$ can also be shown by writing it with a hypergeometric function: $F(x) = \frac13 {}_2F_1(3/2,3/2;5/2;x^2)$. In its series representation, it becomes trivial that $F(x)>0$ for $|x|<1$. Hence, 
\bea
  \frac{\pa G_I}{\pa \rho_I} > 0.
\eea
The unmodified form of $G_I$ is thus a monotonically increasing function of $\rho_I$ in $\wt D^{N_e}$, and so is the modified form in the analytic domain. Note that the shift of $G_I$ by the modification (\ref{def:Gmodification}) maintains this increasing nature.

Next, we show the definiteness of the non-vanishing components of the Jacobian matrix under the condition (\ref{cond:proofs}). The Jacobian $J$ can be written in a unified fashion as below.
\bea
  \frac{\pa G_I}{\pa \rho_J}
  &=&  \delta_{IJ} \frac{\pa G_I}{\pa \rho_I}
\nn&&
   + \delta_{J\in\alpha} \left( 1 - \delta_{IJ} \right)  
     \Biggl[ 
     \frac{(l_J^{(0)})^2}{4} \frac{B_\alpha}{\sqrt{1-\rho_J^2}} 
\nn&&\quad\,\,
       + \frac{l_J^{(0)}}{2 l_I^{(0)}} F(\rho_J) \left\{
          B_\alpha l_I^{(0)} l_J^{(0)} + 4 \delta_{J\in\alpha} \Gamma_\alpha^{(L)} \rho_I \rho_J 
          \right\} \Biggr]
\nn&&
   - \delta_{J\in\beta} \left( 1 - \delta_{IJ} \right)  
     \Biggl[ 
     \frac{(l_J^{(0)})^2}{4} \frac{B_\beta}{\sqrt{1-\rho_J^2}} 
\nn&&\quad\,\,
       + \frac{l_J^{(0)}}{2 l_I^{(0)}} F(\rho_J) \left\{
           B_\beta l_I^{(0)} l_J^{(0)} - 4 \delta_{J\in\beta} \Gamma_\beta^{(L)} \rho_I \rho_J 
          \right\} \Biggr] , 
\nn
\eea
where $\alpha$ and $\beta$ denote the cells to the left and right of the $I$-th edge. The delta function $\delta_{J\in\alpha}$ returns $(\sgn\!{}_J \alpha)$ if the $J$-th edge belongs to the cell $\alpha$ and otherwise vanishes. Therefore, if $(B_\alpha l_I^{(0)} l_J^{(0)} + 4 \Gamma_\alpha^{(L)} \rho_I \rho_J)\geq 0$, then each component of $J$ is definite because the first terms in the brackets are always positive. Since $|\rho_I|\leq 1$, when the condition (\ref{cond:proofs}) holds, the condition for the definiteness also holds. 
\bea
\label{eq:definiteness}
&&  \frac{\pa G_I}{\pa \rho_J} > 0  \quad {\rm for~} J\in \alpha {\rm ~and~}
   (\sgn\!{}_J \alpha) = 1,
\nn
&&  \frac{\pa G_I}{\pa \rho_J} < 0  \quad {\rm for~} J\in \beta {\rm ~and~}
   (\sgn\!{}_J \beta) = 1 . 
\eea
This means that the function $\Vec{G}$ is either constant or monotonic in the analytic domain with respect to any component of $\Vec{\rho}$ under the condition (\ref{cond:proofs}).

Under the same condition, the Jacobian is found to be an invertible matrix. Because no pair of the edges shares the same adjacent cells, no pair of the edges shares the same set of edges that share the same adjacent cells. Consequently, each row vector of $J$, $\frac{\pa G_I}{\pa \Vec{\rho}}$, has different non-zero components from any other row vector $\frac{\pa G_{J\neq I}}{\pa \Vec{\rho}}$ due to the definiteness (\ref{eq:definiteness}). Hence, the row vectors are linearly independent, and $J$ is an $N_e \times N_e$ invertible matrix.

\subsection{A proof for the isolation of the solutions}

Based on the above facts, the solutions of Eq. (\ref{def:vYL}) in the analytic domain are found to all be isolated under the condition (\ref{cond:proofs}). We show its proof by assuming the opposite.   

Suppose a solution is not isolated. This means that there is, at least, one direction in the analytic domain in which Eq. (\ref{def:vYL}) is satisfied. Using the Jacobian matrix $J$, the direction as the vector $\Vec{w} \in D^{N_e}$ must satisfy
\bea
\label{eq:isolation}
  J \Vec{w} = \Vec{0}.
\eea
However, $J$ is an invertible matrix. Therefore, the inverse matrix $J^{-1}$ exists, and $\Vec{w}$ must be zero to satisfy (\ref{eq:isolation}). Hence, there is no such direction around the solution that satisfies Eq. (\ref{def:vYL}), and the solution is isolated, if it exists. Even if a solution is at a boundary, such as at $\rho_I=- 1$, the truncated Jacobian is invertible on the boundary, and so is $J$ in the neighborhood of the solution except for the boundary. It suffices to state that the solution is isolated in the analytic domain.

Note that the above proof shares the same condition for the implicit function theorem and the inverse function theorem.

\subsection{A proof for the uniqueness of the solution}

In the same manner, we can show that the solution in the analytic domain is unique if it exists. We prove this in the same manner as for the isolation, assuming the opposite.

Suppose there are two solutions, $\Vec{\rho}_1$ and $\Vec{\rho}_2$, in the analytic domain. Draw the straight line with the line element, $d\Vec{s}$, so that it connects the two points in the analytic domain. $\Vec{\rho}_1$ and $\Vec{\rho}_2$ are two roots of each $G_I$ on this closed one-dimensional line segment. According to Rolle's theorem, if there are two roots of a function $f$ along a line and $f$ is continuous and differentiable inside the interval of the two roots, there exists a point inside the interval at which the function gives either its maximum or its minimum on the interval and its first derivative vanishes. Provided that each $G_I$ is a continuous and differentiable function on the line between two solutions, the theorem indicates that, for the vector $d\Vec{s}$ parallel to $\Vec{\rho}_2 - \Vec{\rho}_1$, there exists a set of points $\Vec{\varrho}_I$ for $I=1,\ldots,N_e$ on the line segment such that they give the vanishing first derivatives
\bea
\label{eq:Rolles1}
  d\Vec{s} \cdot \nabla G_I(\Vec{\varrho}_I) = 0.
\eea
Because of the definiteness of the first derivative of $G_I$, the collection of the first derivatives at different points constitutes another invertible matrix $\wt J$ analogous to the Jacobian:
\bea
\label{eq:Rolles2}
  d\Vec{s}^T \wt J = \Vec{0},
\eea
where $\wt J_{IJ} \equiv \frac{\pa}{\pa \rho_I} G_J(\Vec{\varrho}_J)$. Because $\wt J$ is invertible, $d\Vec{s}$ must be $\Vec{0}$. Hence, if there is a solution of Eq. (\ref{def:vYL}) under the condition (\ref{cond:proofs}), there is no other solution in the domain that can be reached by the straight line on which $\Vec{G}$ is analytic.

If $V_\alpha$ takes a sufficient value, we can always draw such a line, and the proof holds. However, there are two cases to which the above is not directly applicable. If the two solutions are situated at the same boundary, say $\rho_I=-1$, then $\frac{\pa \Vec{G}}{\pa \rho_I}$ becomes divergent. In such a case, because the $I$-th component of $d\Vec{s}$ is zero, we can neglect the $I$-th contraction of $\wt J$ with $d\Vec{s}$ and follow the same line as the above proof. In this way, we can state that there is only one solution at $\rho_I=-1$ and no other solution in the analytic domain. The other case occurs when $V_\alpha$ is small enough or zero and the two points cannot be connected in a naive way by the straight line on which $\Vec{G}$ is analytic. In other words, a simply drawn straight line between two solutions crosses the border of singularity (\ref{def:domain2}), and $\Vec{G}$ is not analytic on the line. In this case, we have to employ other means to demonstrate the uniqueness.

Although it heavily relies on the existence of the solutions in the analytic domain, there is a straightforward way to show the uniqueness by dragging the zero of the function with a sufficient value of $V_\alpha$. Imagine a sufficient value of $V_\alpha$ such that the unique solution in the analytic domain can reach any point around the boundary of singularity in the $I$-th direction, by the straight line on which $\Vec{G}$ is analytic. Then, we can drag the zero to a different location for a different value of $V_\alpha$ using the analyticity of $\Vec{G}$. Namely, when $V_\alpha$ is changed by $\delta V_\alpha$, the zero is shifted by $\delta \Vec{\rho}$ such that $\Vec{G}\left(\Vec{\rho};V_\alpha\right) = \Vec{G}\left( \Vec{\rho}+\delta\Vec{\rho};V_\alpha+\delta V_\alpha \right)$. The existence of $\delta\Vec{\rho}$ is compensated by the infinitesimal expression of the above. 
\bea
  0 &=& \left(\delta\Vec{\rho} \cdot \nabla \right) 
        \Vec{G}\left(\Vec{\rho};V_\alpha\right)
   + \delta V_\alpha \frac{\pa}{\pa V_\alpha} 
        \Vec{G}\left(\Vec{\rho};V_\alpha\right)
\nn
  0 &=& J\left(\Vec{\rho};V_\alpha\right) \delta\Vec{\rho}
   + \delta V_\alpha \frac{\pa}{\pa V_\alpha} 
        \Vec{G}\left(\Vec{\rho};V_\alpha\right).
\eea
Because $J$ is an invertible matrix, there always exists $\delta\Vec{\rho}$ that satisfies the above relation, regardless of the value of the second term. By the successive applications of the above infinitesimal moves until it reaches $V_\alpha\simeq 0$, we obtain the zero for the case in question. Taking into account the nature of the boundary of singularity, the above analytic flow of the zero cannot go beyond the boundary, and thus the zero is situated inside the boundary of singularity. In addition, because the zero can reach any point around the boundary of singularity by the straight line in the case with a sufficient value of $V_\alpha$, there is no other zero. Thus, the uniqueness of the solution is also assured in the case in question. As such, Proposition \ref{proposition} holds.

\subsection{A few additional remarks on the uniqueness}

The above proofs cannot naively be extended to the special boundary (\ref{def:domain1}, \ref{def:domain2}), and we reserve its proof for a future work. However, its contrapositive statement is proven above and can be proven in a more mathematically elegant way. The statement is as follows: if there is no special boundary, there is one and only one solution in the domain. 

In the absence of a special boundary, the vector field $\Vec{G}$ is always pointing outward at the boundary of the domain, and therefore we can apply the famous Poincar\'e-Hopf theorem, which relates the indices of $\Vec{G}$ with the Euler characteristic:
\bea
  \sum_\gamma {\rm ind}_\gamma \Vec{G} = \chi(\wt D^{N_e}).
\eea
The domain is topologically a $N_e$-dimensional hypercube, and its Euler characteristic is one. Owing to the monotonicity of $G_I$ in the $I$-th direction, the vector $\Vec{G}$ cannot face around any point, and thus there is no sink of $\Vec{G}$. 
Therefore, there exists one and only one solution of $\Vec{G}=\Vec{0}$ as a source of $\Vec{G}$ in the domain.

If another parameterization of the curvature is chosen to describe the model without the saturation bounds, the uniqueness of the solution can be proven as above, or, in general, by the generalized Poincar\'e-Hopf theorem \cite{simsek07} for the vector field $\Vec{G}$ with some incoming flux, provided that such flux is fully investigated.


\end{document}